\documentclass{aa} 

\usepackage{psfig}
\usepackage{graphicx}

\begin{document}

\title{Stability of hydrodynamical relativistic planar jets. \\ 
  II. Long-term nonlinear evolution}

\subtitle{}

\author{M. Perucho\inst{1}
          \and
	J. M. Mart\'{\i}\inst{1}
          \and
	M. Hanasz\inst{2}
}

   \offprints{M. Perucho}

   \institute{Departamento de Astronom\'{\i}a y Astrof\'{\i}sica,
              Universidad de Valencia, 46100 Burjassot (SPAIN)\\ 
   \email{manuel.perucho@uv.es,
              jose-maria.marti@uv.es} \and Toru\'n Centre for
              Astronomy, Nicholas Copernicus University, PL-97-148
              Piwnice k.Torunia (POLAND)\\ 
   \email{mhanasz@astri.uni.torun.pl}}

\date{Received .../ Accepted ...}

\abstract{In this paper we continue our study of the Kelvin-Helmholtz (KH)
instability in relativistic planar jets following the long-term
evolution of the numerical simulations which were introduced
in Paper I. The models have been classified into four classes (I to
IV) with regard to their evolution in the nonlinear phase,
characterized by the process of jet/ambient mixing and momentum
transfer. Models undergoing qualitatively different non-linear
evolution are clearly grouped in well-separated regions in a jet
Lorentz factor/jet-to-ambient enthalpy diagram. Jets with a low Lorentz
factor and small enthalpy ratio are disrupted by a strong shock after
saturation. Those with a large Lorentz factor and enthalpy ratio are
unstable although the process of mixing and momentum exchange proceeds
to a longer time scale due to a steady conversion of kinetic
to internal energy in the jet. In these cases, the high value of the
initial Lorentz seems to prevent transversal velocity from growing far
enough to generate the strong shock that breaks the slower jets.
Finally, jets with either high Lorentz factors and small enthalpy
ratios or low Lorentz factors and large enthalpy ratios appear as the
most stable.

  In the long term, all the models develop a distinct transversal
structure (shear/transition layers) as a consequence of KH
perturbation growth. The properties of these shear layers are analyzed
in connection with the parameters of the original jet models.

\keywords{Galaxies: jets - hydrodynamics - instabilities} }

\titlerunning{}
\authorrunning{M. Perucho et al.}
\titlerunning{Stability of hydrodynamical relativistic planar jets. II}

\maketitle

\section{Introduction}

  In the previous paper (Perucho et al. 2004), hereafter
as Paper I, we have performed several simulations and characterized
the effects of relativistic dynamics and thermodynamics in the
development of KH instabilities in planar, relativistic jets. We
performed a linear stability analysis and numerical simulations for the
most unstable first reflection modes in the temporal approach, for
three different values of the Lorentz factor $\gamma$ (5, 10 and 20)
and a few different values of specific internal energy of the jet
matter (from $0.08$ to $60.0 c^2$; $c$ is the light speed in vacuum).

  In Paper I we focused on the linear and postlinear regime,
especially on the comparison of the results of the linear stability
analysis and numerical simulations in the linear range of KH
instability.

  We demonstrated first that, with the appropriate numerical
resolution, a high convergence could be reached between the growth
rate of perturbed modes in the simulations and the results of a linear
stability analysis performed in the vortex sheet approximation. A 20\%
accuracy has been determined for the full set of simulated models, on
average.  The agreement between the linear stability analysis and
numerical simulations of KH instability in the linear range has been
achieved for a very high radial resolution of 400 zones/$R_j$ ($R_j$
is the jet radius), which
appears to be especially relevant for hot jets. The verified high
accuracy of the simulations made it possible to extend the analysis up
to nonlinear phases of the evolution of KH instability. We identified
several phases in the evolution of all the models: linear, saturation
and mixing phases. The further analysis of the mixing process,
operating during the long-term evolution, will be performed in the
present paper.

  We have found that in each of the examined cases the linear phase
always ends when the longitudinal velocity perturbation departs from
linear growth. Then the longitudinal velocity saturates at a value
close to the speed of light. This limitation (inherent to relativistic
dynamics) is easily noticeable in the jet reference frame. We also
noted a saturation of the transversal velocity perturbation, measured
in the jet reference frame, at the level of about $0.5c$. The
transversal velocity saturates later than the longitudinal velocity at
a moment which is close to the peak value of pressure perturbation.

  Therefore we concluded that the relativistic nature of the examined
flow is responsible for the departure of the system from linear
evolution, which manifests as the limitation of velocity
components.  This behaviour is consistent with the one deduced by
Hanasz (1995, 1997) with the aid of analytical methods.

  Our simulations, performed for the most unstable first reflection
modes, confirm the general trends resulting from the linear stability
analysis: the faster (larger Lorentz factor) and colder jets have
smaller growth rates.  As we mentioned in the Introduction of Paper I,
Hardee et al. (1998) and Rosen et al. (1999) note an exception which
occurs for the hottest jets. These jets appear to be the most stable
in their simulations (see also the simulations in Mart\'{\i} et
al. 1997). They suggest that this behaviour is caused by the lack of
appropriate perturbations to couple to the unstable modes. This could
be partially true as fast, hot jets do not generate overpressured
cocoons that let the jet run directly into the nonlinear regime.
However, from the point of view of our results in Paper I, the high
stability of hot jets may have been caused by the lack of radial
resolution that leads to a damping in the perturbation growth
rates. Finally, one should keep in mind that the simulations performed
in the aforementioned papers only covered about one hundred time units
($R_j/c$), well inside the linear regime of the corresponding models
for small perturbations. In this paper, the problem of the stability
of relativistic cold, hot, slow and fast jets is analyzed on the basis
of long-term simulations extending over the fully nonlinear evolution
of KH instabilities.

  Finally, we found in Paper I that the structure of the
pertubation does not change much accross the linear and saturation
phases, except that the oblique sound waves forming the perturbation
became steep due to their large amplitude. In this paper we 
show that the similarity of all models at the saturation time,
found in Paper I, will lead to different final states in the course of the
fully nonlinear evolution.

  The plan of this paper is as follows: In Section~\ref{sect:numsim}
we recall the parameters of the simulations discussed in Paper I and
in this paper along with a new complementary set of simulations
introduced here with the aim of making the analysis more
general. In Section~\ref{sect:results} we describe our new results
concerning the long-term nonlinear evolution of KH modes and 
we discuss our results and conclude in
Section~\ref{sect:concl}.

\section{Numerical simulations \label{sect:numsim}} 

  In this paper we continue the analysis of simulations presented in
Paper I. The parameters of the simulations are listed in
Table~\ref{tab:param}. The values of the parameters
were chosen to be close to those used in some simulations by Hardee et
al. (1998) and Rosen et al. (1999) and to span a wide range in
thermodynamical properties as well as jet flow Lorentz factors. In all
the simulations of Paper I, the density in the jet and ambient gases
are $\rho_{0j}=0.1$, $\rho_{0a}=1$ respectively and the adiabatic
exponent $\Gamma_{j,a}=4/3$.

  Since the internal rest mass density is fixed, there are two free
parameters characterizing the jet equilibrium: Lorentz factor and jet
specific internal energy displayed in columns 2 and 3 of
Table~\ref{tab:param}. Models whose names start with the same letter
have the same thermodynamical properties. Jet (and ambient) specific
internal energies grow from models A to D. Three different values of
the jet flow Lorentz factor have been considered for models B, C and
D. The other dependent parameters are displayed in columns 5-11 of
Table~\ref{tab:param}. Note that given our choice of $\rho_{0j}$, the
ambient media associated with hotter models are also hotter. The next
three columns show the longitudinal wavenumber together with
oscillation frequency and the growth rate of the most unstable first
reflection (body) mode. The following three columns display the same
quantities in the the jet reference frame. The next two columns show the
transversal wavenumbers of linear sound waves in jet and ambient
medium respectively. The last column shows the linear growth rate of
KH instability in the jet reference frame expressed in dynamical time
units, i.e. in which time is scaled to $R_j/c_{sj}$.  All other
quantities in the table are expressed in units of the ambient
density, $\rho_{0a}$, the speed of light, $c$, and the jet radius,
$R_j$.

  In order to extend our conclusions to a wider region in the initial
parameter space, we have performed a new set of simulations, which
will be discussed only in some selected aspects. The initial data for
these new simulations are compiled and shown in the lower part of
Table \ref{tab:param}.  The external medium in all cases is that of
model A05. From models F to H, internal energy in the jet is increased
and rest-mass density decreased in order to keep pressure equilibrium,
whereas the jet Lorentz factor is kept equal to its value in model
A05.

  The initial momentum density in the jet decreases along the
sequence A05, F, G, H. Simulations I and J have the same
thermodynamical values as models F and G, respectively, but have
increasing jet Lorentz factors to keep the same initial momentum
density as model A05. Finally, in simulations K and L we exchange the
values of the Lorentz factor with respect to those in runs I and J.

\begin{table*}
\begin{center}
$
\begin{array}{|c|ccc|ccccccc|ccc|ccc|ccc|}
\hline
 {\rm Model} &        \gamma& \rho_{0j} &    \varepsilon_j &    \varepsilon_a  &     c_{sj}  &        c_{sa} &            
    p &           \nu &          \eta &           M_j & k_{\parallel} &      \omega_r &      \omega_i  
       &{k'_\parallel} &   \omega'_{r} & \omega'_i   & k_{j\perp} &   
       k_{a\perp} & {\omega'}^{\rm dyn}_i\\
\hline
         {\ rm A05} &            5 &  0.1&        0.08 &          0.008 &    0.18 &          0.059 &          0.0027 &            0.11 &            0.11 &            5.47 &            0.30 &            0.20 &          0.026 &            1.32 &            7.20 &            0.13 &            7.08 &            0.53 &            0.73 \\
         {\rm B05} &            5 &  0.1&        0.42 &          0.042 &    0.35 &          0.133 &          0.014 &            0.14 &            0.15 &            2.83 &            0.69 &            0.49 &          0.055 &            2.62 &            7.32 &            0.28 &            6.84 &            1.08 &            0.79 \\
         {\rm C05} &            5 &  0.1&        6.14 &          0.614 &    0.55 &	  0.387 &          0.205 &            0.44 &            0.51 &            1.80 &            2.00 &            1.60 &          0.114 &            5.73 &            9.98 &            0.57 &            8.17 &            1.07 &            1.05 \\
         {\rm D05} &            5 &  0.1&        60.0 &          6.000 &    0.57 &          0.544 &          2.000 &            0.87 &            0.90 &            1.71 &            2.63 &            2.18 &          0.132 &            7.02 &           11.56 &            0.66 &            9.18 &            0.24 &            1.15 \\
         {\rm B10} &           10 &  0.1&        0.42 &          0.042 &    0.35 &          0.133 &          0.014 &            0.14 &            0.15 &            2.88 &            0.50 &            0.41 &          0.031 &            3.59 &           10.28 &            0.31 &            9.64 &            0.94 &            0.90 \\
         {\rm C10} &           10 &  0.1&        6.14 &          0.614 &    0.55 &          0.387 &          0.205 &            0.44 &            0.51 &            1.83 &            1.91 &            1.72 &          0.055 &            9.77 &           17.67 &            0.55 &           14.72 &            1.49 &            1.01 \\
         {\rm D10} &           10 &  0.1&        60.0 &          6.000 &    0.57 &          0.544 &          2.000 &            0.87 &            0.90 &            1.73 &            2.00 &            1.81 &          0.063 &            9.67 &           16.58 &            0.63 &           13.47 &            0.20 &            1.10 \\
         {\rm B20} &           20 &  0.1&        0.42 &          0.042 &    0.35 &          0.133 &          0.014 &            0.14 &            0.15 &            2.89 &            0.46 &            0.39 &          0.014 &            6.51 &           18.76 &            0.28 &           17.60 &            0.90 &            0.81 \\
         {\rm C20} &           20 &  0.1&        6.14 &          0.614 &    0.55 &        0.387 &          0.205 &            0.44 &            0.51 &            1.83 &            1.44 &            1.37 &          0.027 &           13.89 &           25.38 &            0.54 &           21.24 &            1.28 &            0.99 \\
         {\rm D20} &           20 &  0.1&        60.0 &          6.000 &    0.57 &          0.544 &          2.000 &            0.87 &            0.90 &            1.74 &            2.00 &            1.91 &          0.029 &           18.11 &           31.43 &            0.58 &           25.68 &            0.31 &            1.01 \\
\hline
           {\rm F} &          5 &  0.01 &          0.77 &         0.008 &           0.41 &          0.058 &          0.0026 &        0.018 &            0.02 &            2.38 &            0.46 &            0.53 &          0.14 &            1.23 &            2.83 &            0.70 &            2.55 &            3.72 &            1.70 \\
           {\rm G} &          5 & 0.001 &          7.65 &         0.008 &           0.55 &          0.058 &          0.0026 &        0.009 &            0.01 &            1.78 &            0.66 &            0.53 &          0.15 &            1.87 &            3.22 &            0.75 &            2.62 &            4.99 &            1.36 \\
           {\rm H} &          5 &0.0001 &          76.5 &         0.008 &           0.57 &          0.058 &          0.0026 &        0.008 &            0.01 &            1.71 &            0.66 &            0.48 &          0.15 &            1.95 &            3.23 &            0.75 &            2.57 &            4.71 &            1.31 \\
           {\rm I} &       11.7 &  0.01 &          0.77 &         0.008 &           0.41 &          0.058 &          0.0026 &         0.018 &            0.02 &            2.42 &            0.30 &            0.50 &          0.07 &            1.11 &            2.66 &            0.82 &            2.42 &            3.52 &            1.99 \\
           {\rm J} &       15.7 & 0.001 &          7.65 &         0.008 &           0.55 &          0.058 &          0.0026 &         0.009 &            0.01 &            1.81 &            0.35 &            0.44 &          0.058 &            1.70 &            3.05 &            0.91 &            2.53 &            4.16 &            1.65 \\
           {\rm K} &       15.7 &  0.01 &          0.77 &         0.008 &           0.41 &          0.058 &          0.0026 &         0.018 &            0.02 &            2.43 &            0.30 &            0.55 &          0.054 &            1.17 &            2.80 &            0.85 &            2.55 &            3.87 &            2.06 \\
           {\rm L} &       11.7 & 0.001 &          7.65 &         0.008 &           0.55 &          0.058 &          0.0026 &         0.009 &            0.01 &            1.81 &            0.30 &            0.31 &          0.069 &            1.52 &            2.72 &            0.81 &            2.26 &            2.93 &            1.47 \\
\hline
\end{array}
$
\end{center}

\caption{Equilibrium parameters of different simulated jet models
along with solutions of the dispersion relation (23) in Paper I,
corresponding to fastest growing first reflection mode, taken as input
parameters for numerical simulations.  The primes are used to assign
wavenumber and complex frequency in the reference frame comoving with
jet. The listed equilibrium parameters are: $\gamma$ - jet Lorentz
factor, $\rho_{0j}$ - rest mass density, $\varepsilon_j$ and
$\varepsilon_a$ - specific internal energies of jet and ambient
medium, $c_{sj}$, $c_{sa}$ - the sound speeds in jet and ambient
medium, $p$ - pressure, $\nu$, $\eta$ - relativistic density and
enthalpy contrasts and $M_j$ - the jet Mach number. All the quantities
in the table, except the last column, are expressed in units of the
ambient density, $\rho_{0a}$, the speed of light, $c$, and the jet
radius, $R_j$. Parameters for the new set of simulations are shown in
the lower part of the table. }

\label{tab:param}    
\end{table*}

The numerical setup for the simulations described in this paper is the
same as in Paper I. We only simulate half of the jet
($x>0$) due to the assumed symmetry of perturbations. Reflecting
boundary conditions are imposed on the symmetry plane of the flow,
whereas periodical conditions are settled on both upstream and
downstream boundaries.  The applied resolution of $400 \times 16$ grid
zones per jet radius is chosen for all simulations listed in
Table~\ref{tab:param}, following the tests described in the Appendix
of Paper I. In the Appendix of this paper we present a discussion of
the influence of different longitudinal resolutions on the simulation
results during the long-term nonlinear evolution. Other details of the
numerical setup are described in Paper I.

  The steady model is then perturbed according to the selected mode
(the most unstable first reflection mode), with an absolute value of
the pressure perturbation amplitude inside the jet of
$p_j^{\pm}=10^{-5}$. This means that those models with the smallest
pressure, like model A, have relative perturbations in pressure three
orders of magnitude larger than those with the highest pressures,
D. However this difference seems not to affect the linear and
postlinear evolution (see footnote in Sect.~\ref{sect:results}).

\section{Results \label{sect:results}}

  Following the behaviour of simulated models, we found in Paper I that
the evolution of the perturbations can be divided into the linear phase,
saturation phase and mixing phase. This section is devoted to
describing the fully nonlinear evolution of the modes described in
Paper I. Our description shares many points with the framework
developed by Bodo et al. (1994) for the case of classical jets.

  In order to illustrate the growth of perturbations and determine the
duration of the linear and saturation phases in our simulations, we
plotted in Fig.~1 of Paper I the amplitudes of the perturbations of
the longitudinal and transversal velocities inside the jet and in the
jet reference frame, together with the pressure oscillation
amplitude. We also plotted the growth of the imposed eigenmodes
resulting from the linear stability analysis. Both velocity
perturbations are transformed from the ambient rest frame to the
unperturbed jet rest frame using the Lorentz transformation rules for
velocity components.

  In Paper I we defined the characteristic times $t_{\rm lin}$ and
$t_{\rm sat}$ as the end of the linear and saturation phases
respectively. At the saturation time, the perturbation structure is
still close to the structure of the initial perturbation, except that
the oblique sound waves forming the perturbation became steeper, 
leading to the formation of shock waves. Finally, nearly all the
simulations lead to a sharp peak of the pressure oscillation
amplitude. The times at which this peak appears, $t_{\rm peak}$, are
equal to or slightly larger than the saturation times for different
models. The definition of $t_{\rm lin}$, $t_{\rm sat}$ and $t_{\rm
peak}$ has been illustrated in Fig.~1 of paper
I. Table~\ref{tab:phases} collects the times of the linear and
saturation phases in the different models (columns 2-4) along with
other characteristic times (defined in the table caption and in
Subsection~\ref{ss:mixdis}).

\begin{table}
$
\begin{array}{ccccccccc}
\hline {\rm Model} & t_{\rm lin} & t_{\rm mex} & t_{\rm mix} & t_{\rm
sat} & t_{\rm peak} & \Delta_{\rm peak} & t_{\rm fmix} & t_{\rm meq}
\\ \hline {\rm A05} & 180 & 335 & 335 & 380 & 380 & 100 & 380 & - \\
{\rm B05} & 125 & 175 & 185 & 200 & 205 & 70 & 210 & 215 \\ {\rm C05}
& 100 & 115 & 120 & 125 & 130 & 5 & >595 & 195 \\ {\rm D05} & 105 &
115 & 115 & 120 & 130 & 5 & >595 & 185 \\ \hline {\rm B10} & 235 & 375
& 335 & 380 & 385 & 100 & 445 & - \\ {\rm C10} & 210 & 245 & 240 & 245
& 250 & 10 & >595 & - \\ {\rm D10} & 180 & 220 & 215 & 225 & 225 & 10
& 350 & 345 \\ \hline {\rm B20} & 450 & 775 & 625 & 760 & 780 & 100 &
>1000 & - \\ {\rm C20} & 270 & 675 & 595 & 645 & 775 & 5 & >1000 &
>1000\\ {\rm D20} & 350 & 465 & 450 & 480 & 500 & 10 & >1000 & >1000\\
\hline
\end{array}
$
\caption{ Times for the different phases in the evolution of the
perturbed jet models. $t_{\rm lin}$: end of linear phase (the
amplitudes of the different quantities are not constant any
longer). $t_{\rm sat}$: end of saturation phase (the amplitude of the
transverse speed perturbation reaches its maximum). $t_{\rm mix}$: the
tracer starts to spread.  $t_{\rm peak}$: the peak in the amplitude of
the pressure perturbation is reached. $t_{\rm fmix}$: external
material reaches the jet axis. $t_{\rm mex}$: the jet has transferred
to the ambient a 1\% of its initial momentum. $t_{\rm meq}$:
longitudinal momentum in the jet and the ambient reach
equipartition. $\Delta_{\rm peak}$: relative value of pressure
oscillation amplitude at the peak (see Fig.~1 of Paper I). Note that,
as a general trend, $t_{\rm lin} < t_{\rm mex} \approx t_{\rm mix}
<t_{\rm sat} <t_{\rm peak} <t_{\rm fmix} < t_{\rm meq}$.  }
\label{tab:phases}   

\end{table}

\begin{figure*} 
\centerline{
\psfig{file=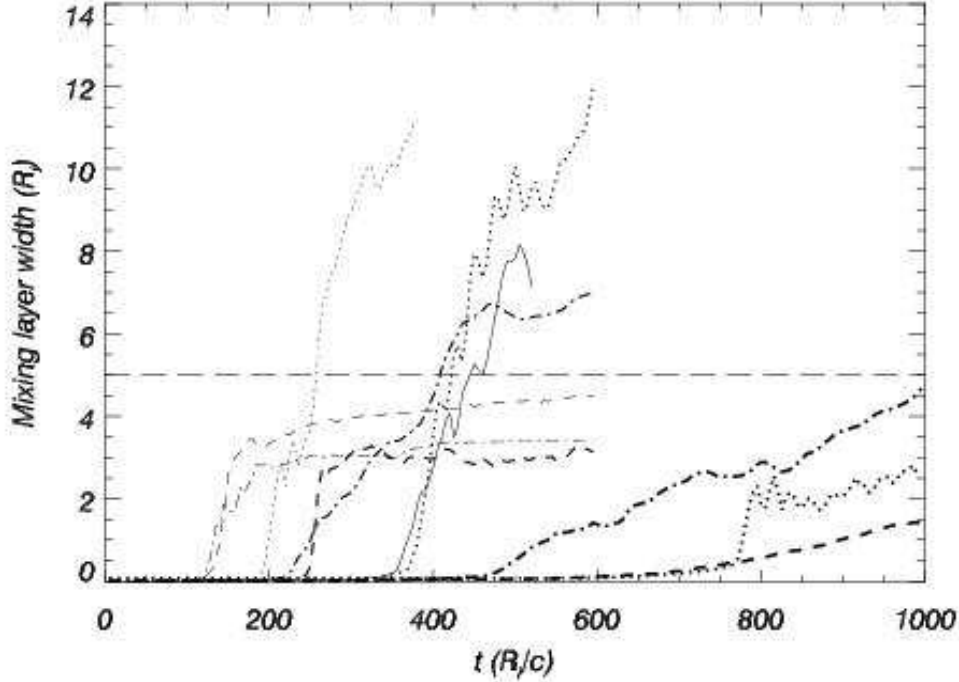,width=0.8\textwidth,angle=0,clip=0} 
}

\caption{Evolution of the mean width of the jet/ambient mixing layer
  with time for all the simulations. Different types of lines are used
  for models with different internal energies: Continuous line: model
  A; dotted line: model B; dashed line: model C; dashed-dotted line:
  model D. Line thickness increases with Lorentz factor (from 5,
  thinest line, to 20 thickest one). A value of 5 $R_{\rm j}$ for the
  width of the mixing layer (horizontal dashed line) serves to
  classify the evolution of the different models.}
       
\label{fig:tracerdisp}
\end{figure*}      

\subsection{Fully nonlinear evolution: jet/ambient mixing
\label{ss:mixdis}}

  The beginning of the mixing phase can be detected by the spreading
of the tracer contours. This can be seen in Fig.~\ref{fig:tracerdisp},
where the evolution with time of the mean width of the layer with
tracer values between 0.05 and 0.95 is shown. The times at which the
mixing phase starts ($t_{\rm mix}$) are shown in
Table~\ref{tab:phases}.  Consistently with the width of the initial
shear layer in our simulations (around 0.1 $R_{\rm j}$), we have
defined $t_{\rm mix}$ as the time at which the mixing layer exceeds a
width of 0.1 $R_{\rm j}$.

  For models with the same thermodynamical properties, those with
smaller Lorentz factors start to mix earlier (see
Table~\ref{tab:phases}). Moreover, according to
Fig.~\ref{fig:tracerdisp}, the models can be sampled in two (or perhaps 
three) categories. Models A05, B05, B10 and D10 have wide ($>5 R_{\rm
j}$) shear layers which are still in a process of widening at the end
of the simulation. The rest of the models have thiner shear layers
($<5 R_{\rm j}$ wide) which are inflating at smaller speeds ($0.5-2
\,\, 10^{-3}c$, in the case of models C05, D05 and C10; $0.5-1.2 \,\,
10^{-2} c$, in the case of models B20, C20, and D20).  A deeper
analysis of the process of widening of the shear layers as a function
of time shows that all the models undergo a phase of exponential
growth extending from $t_{\rm mix}$ to soon after $t_{\rm peak}$.

  We note that those models developing wider mixing layers are those
in which the peak in the maxima of the pressure perturbation as a
function of time, $\Delta_{\rm peak}$, reach values of the order of
$70-100$, with the exception of model B20 that has a thin mixing layer
at the end of the simulation but has $\Delta_{\rm peak} \approx 100$,
and model D10, which does develop a wide shear layer but for which
$\Delta_{\rm peak}$ remains small. We also note that (with the
exception of model D10) the models developing wide mixing layers are
those with smaller internal energies and also relatively smaller
Lorentz factors.

  There are two basic mechanisms that contribute to the process of
mixing between ambient and jet materials. The first one is the
deformation of the jet surface by large amplitude waves during the
saturation phase. This deformation favors the transfer of momentum
from the jet to the ambient medium and, at the same time, the
entrainment of ambient material in the jet. From
Table~\ref{tab:phases}, it is seen that the process of mixing and
momentum exchange overlap during the saturation phase.

\begin{figure*} 
\centerline{ \psfig{file=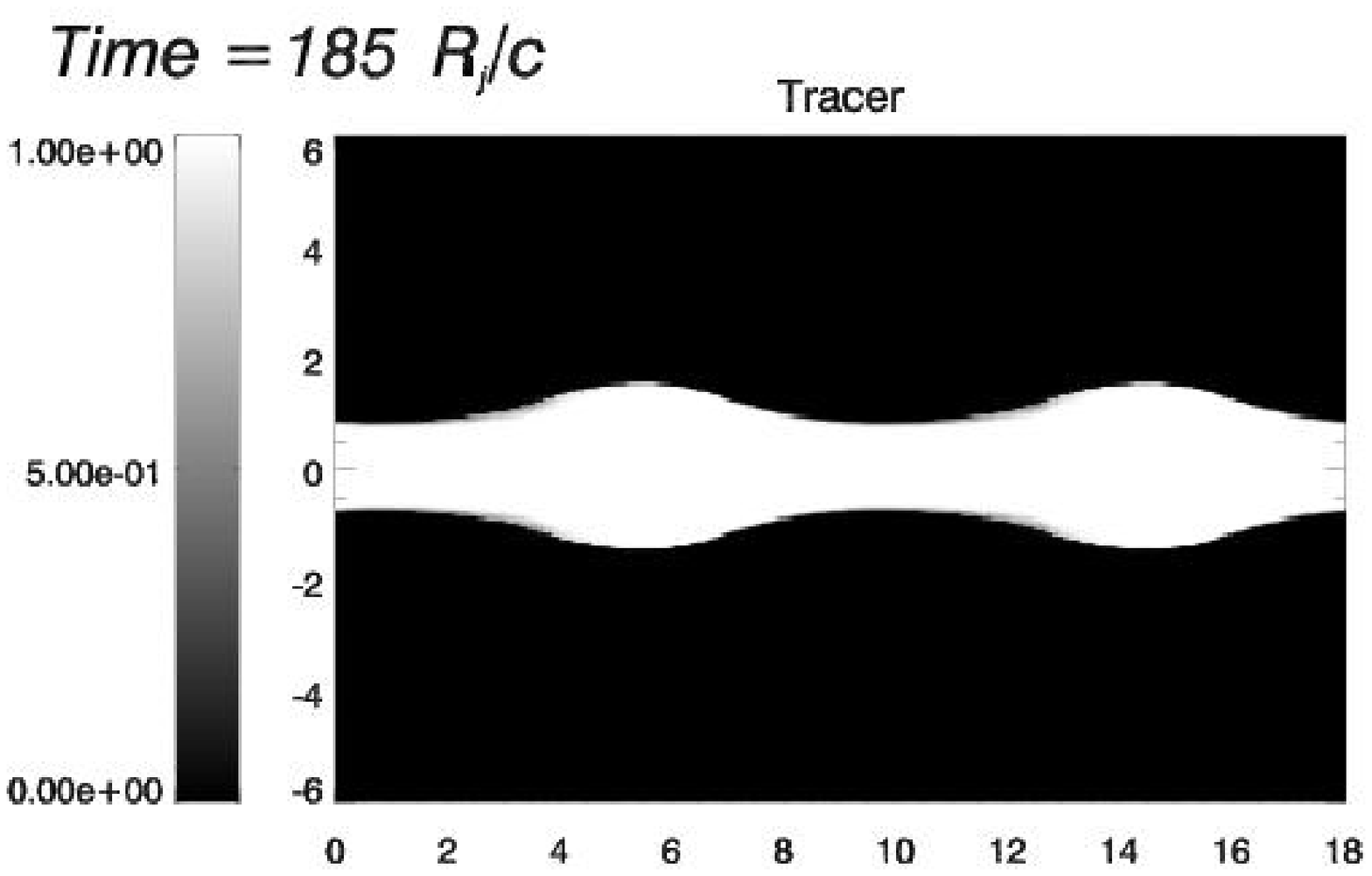,width=0.40
\textwidth,angle=0,clip=} \quad
\psfig{file=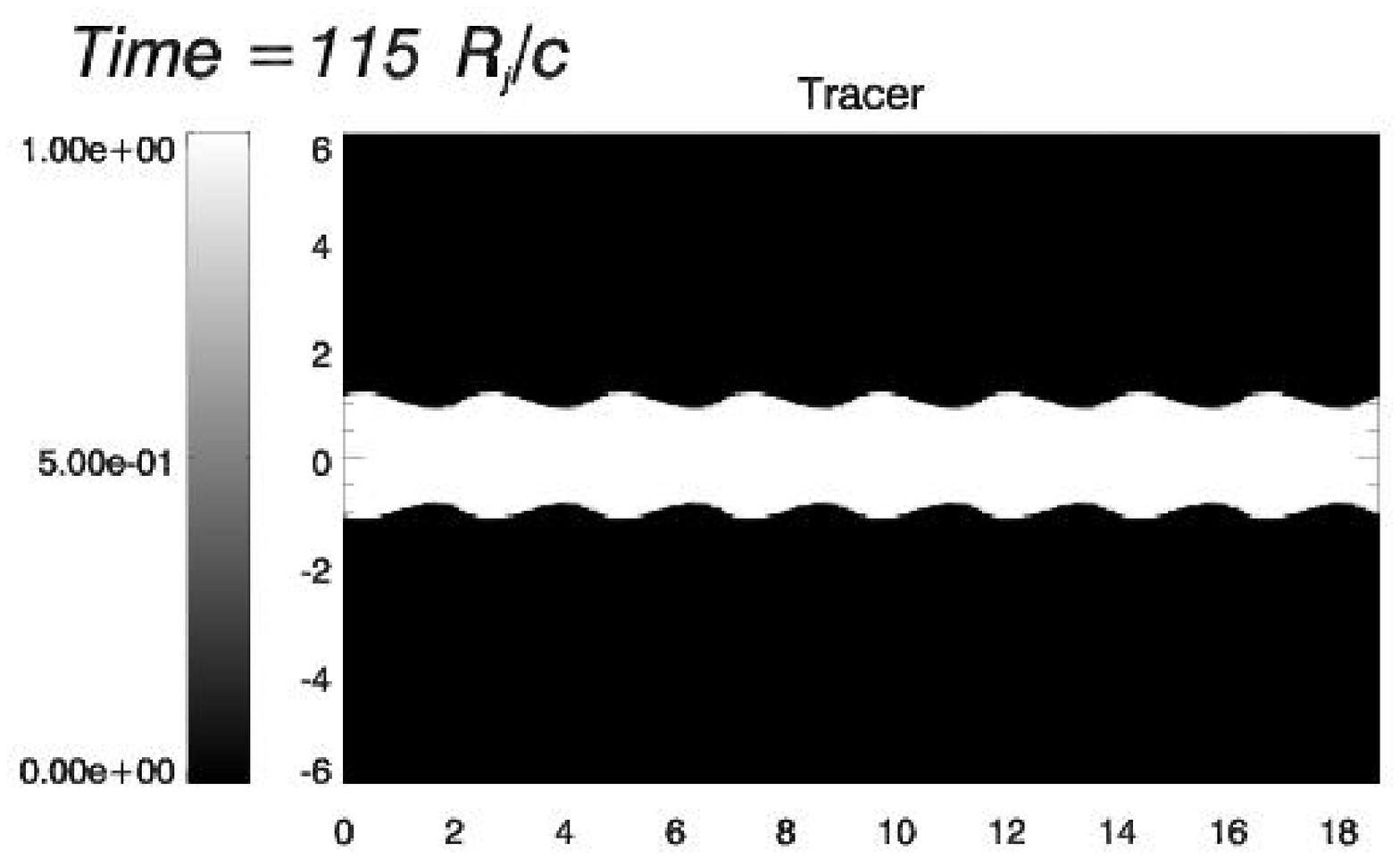,width=0.40 \textwidth,angle=0,clip=} }
\centerline{ \psfig{file=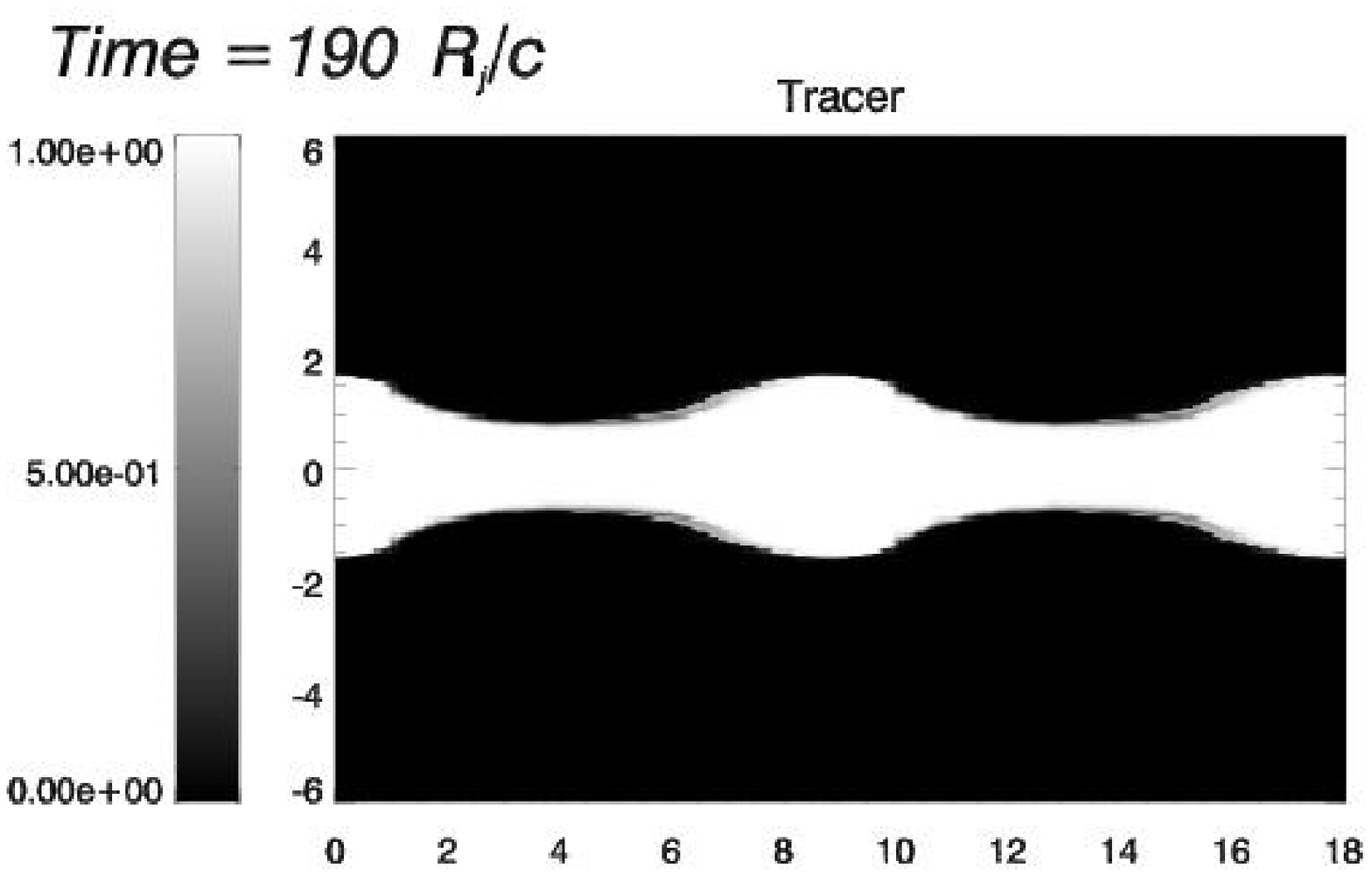,width=0.40
\textwidth,angle=0,clip=} \quad
\psfig{file=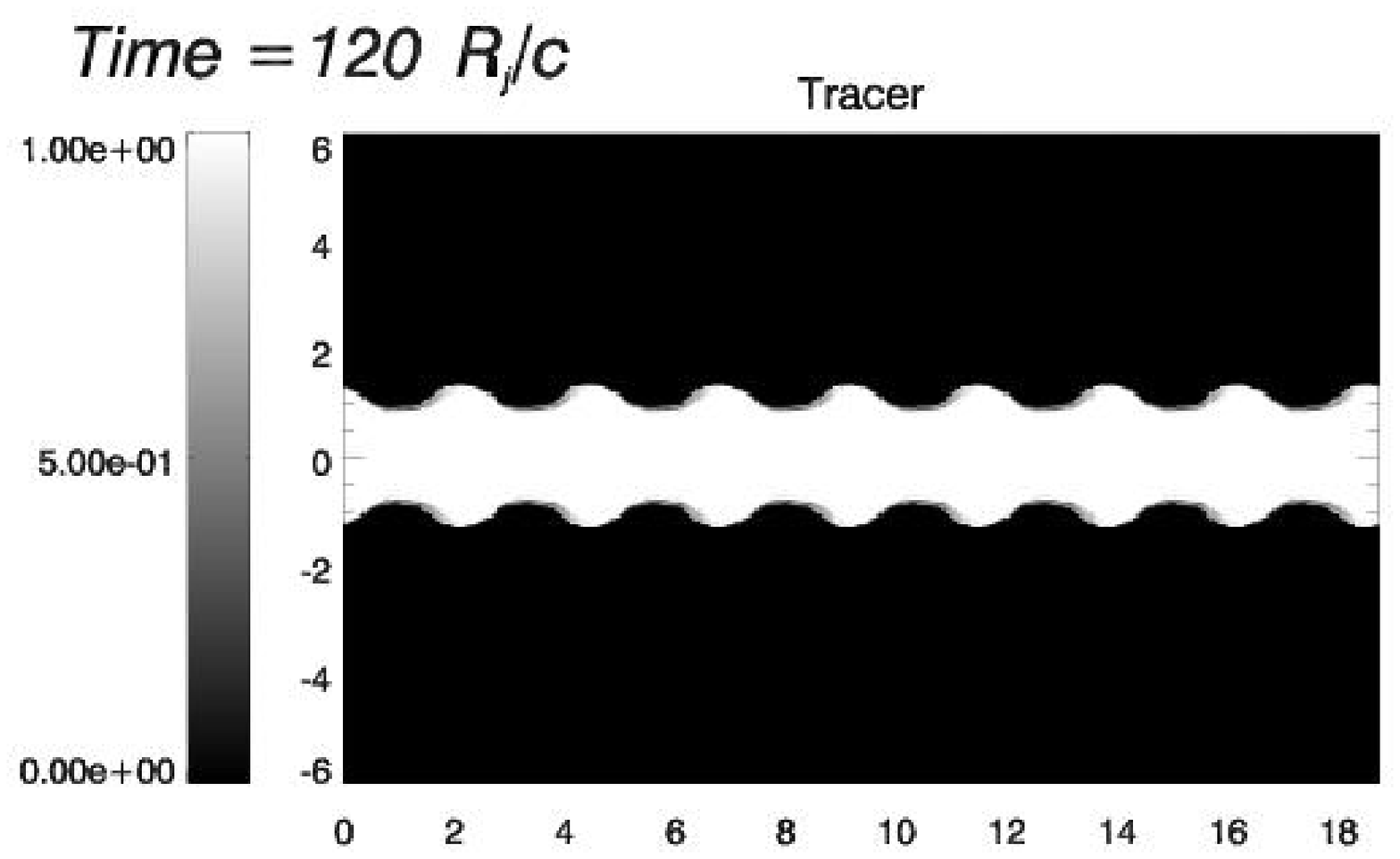,width=0.40 \textwidth,angle=0,clip=} }
\centerline{ \psfig{file=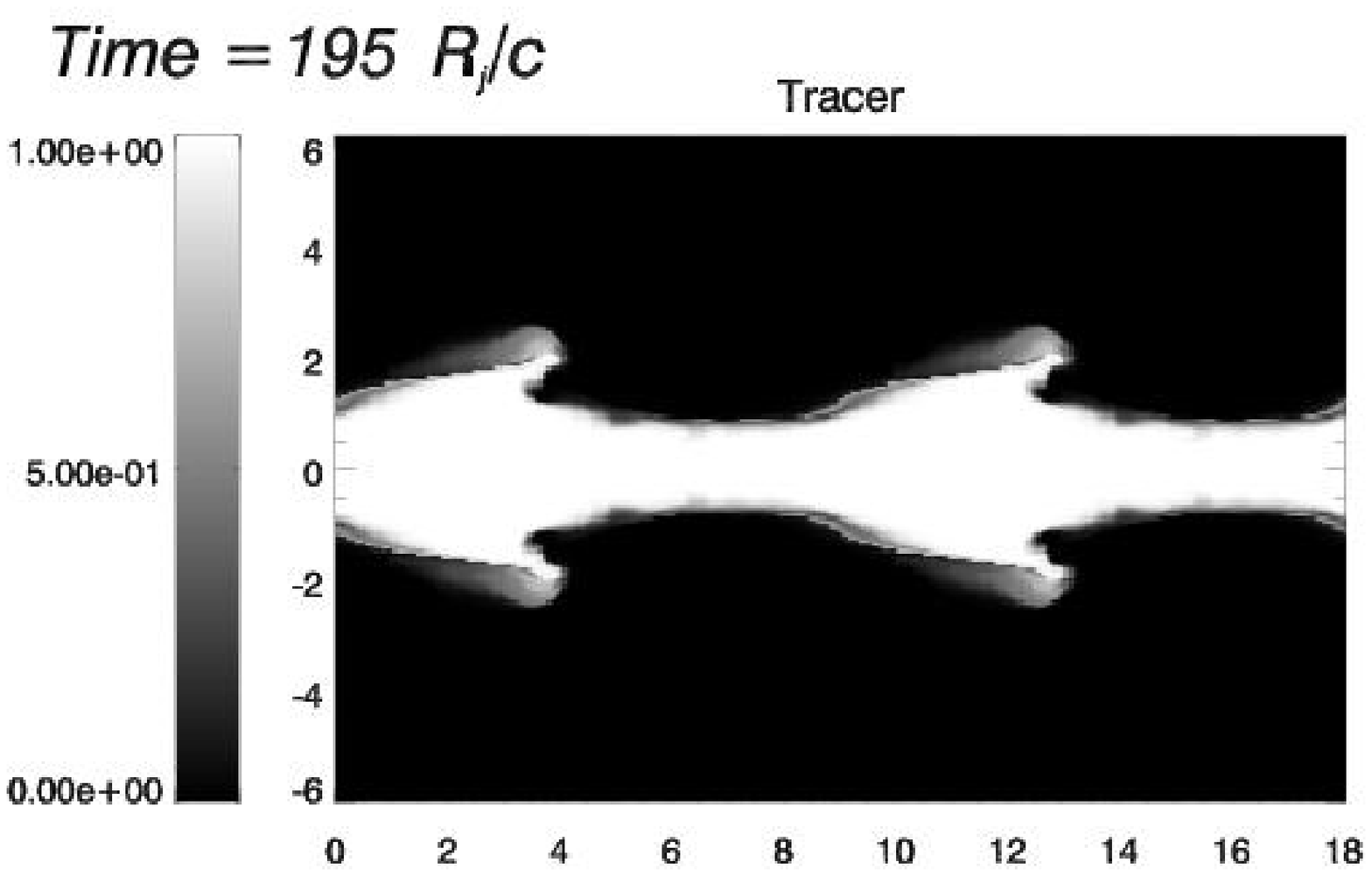,width=0.40
\textwidth,angle=0,clip=} \quad
\psfig{file=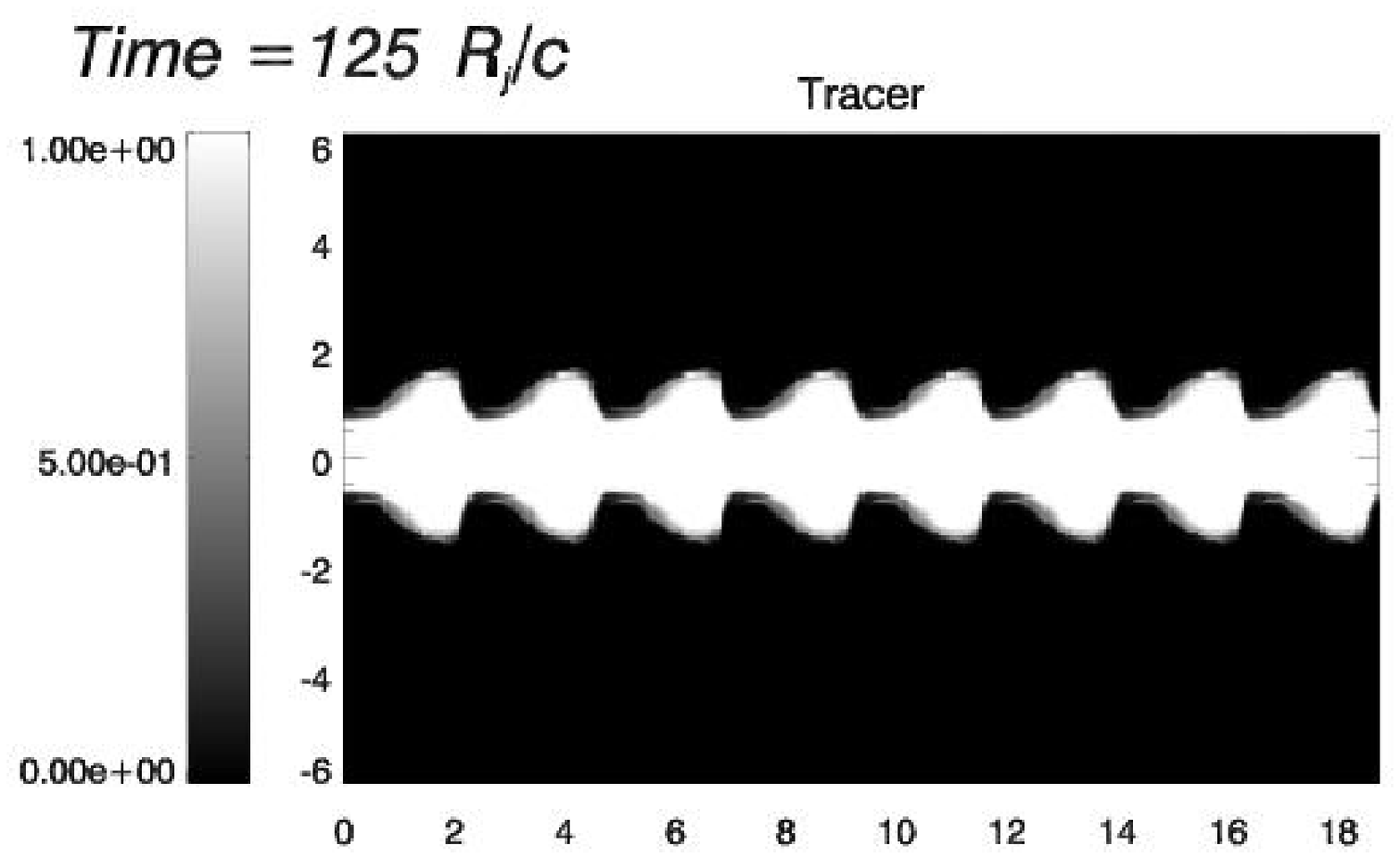,width=0.40 \textwidth,angle=0,clip=} }
\centerline{ \psfig{file=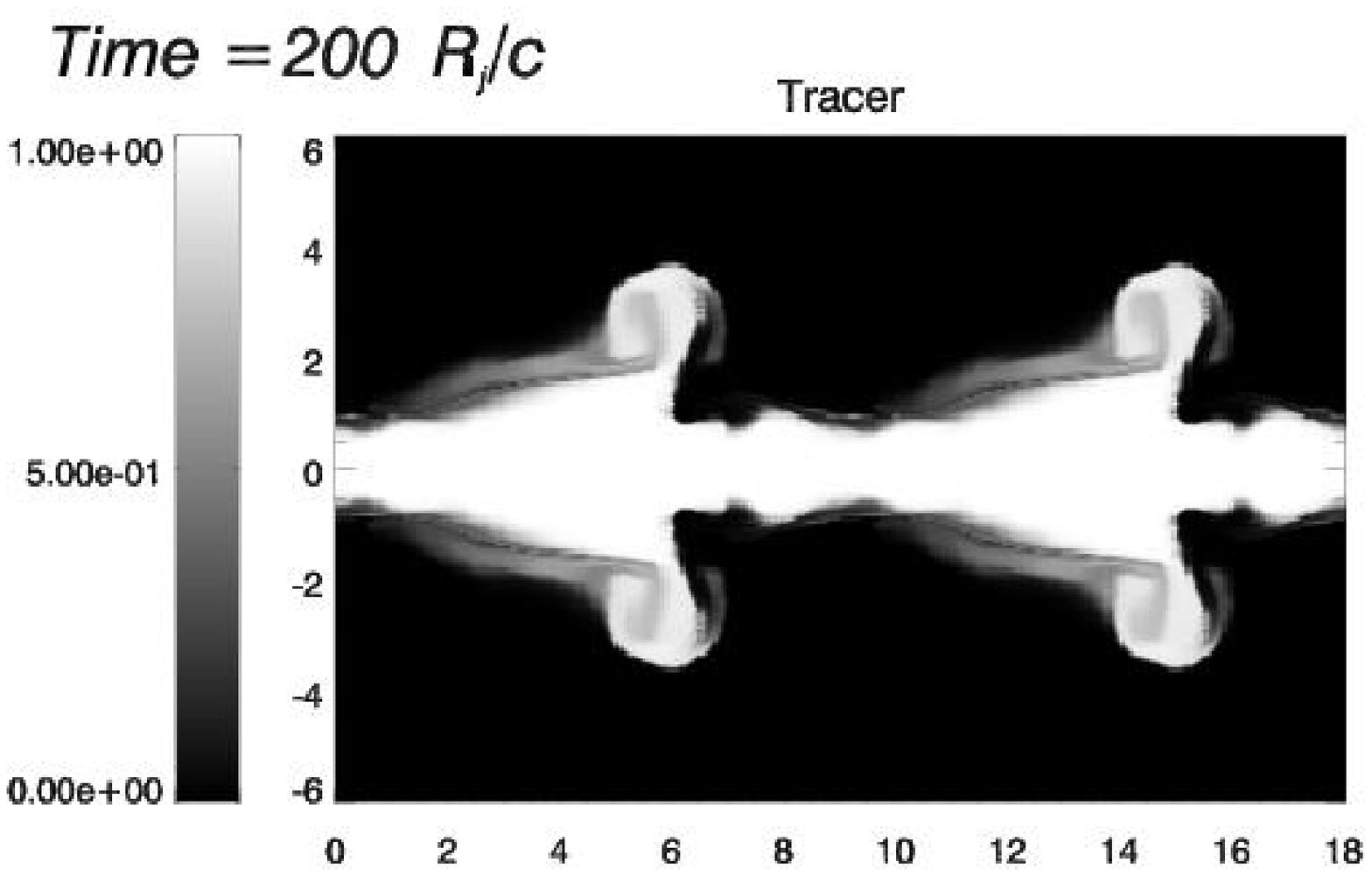,width=0.40
\textwidth,angle=0,clip=} \quad
\psfig{file=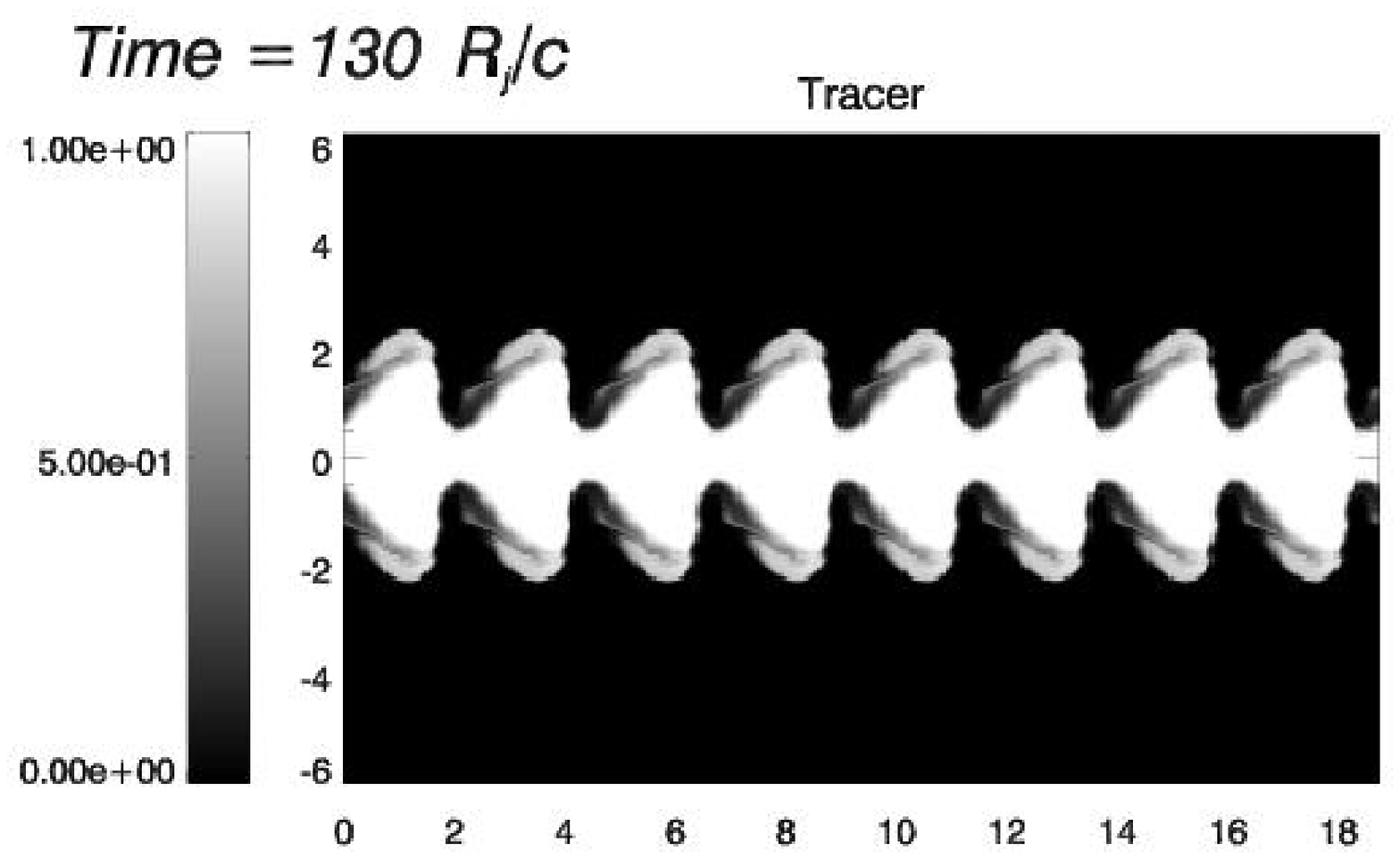,width=0.40 \textwidth,angle=0,clip=} }
\centerline{ \psfig{file=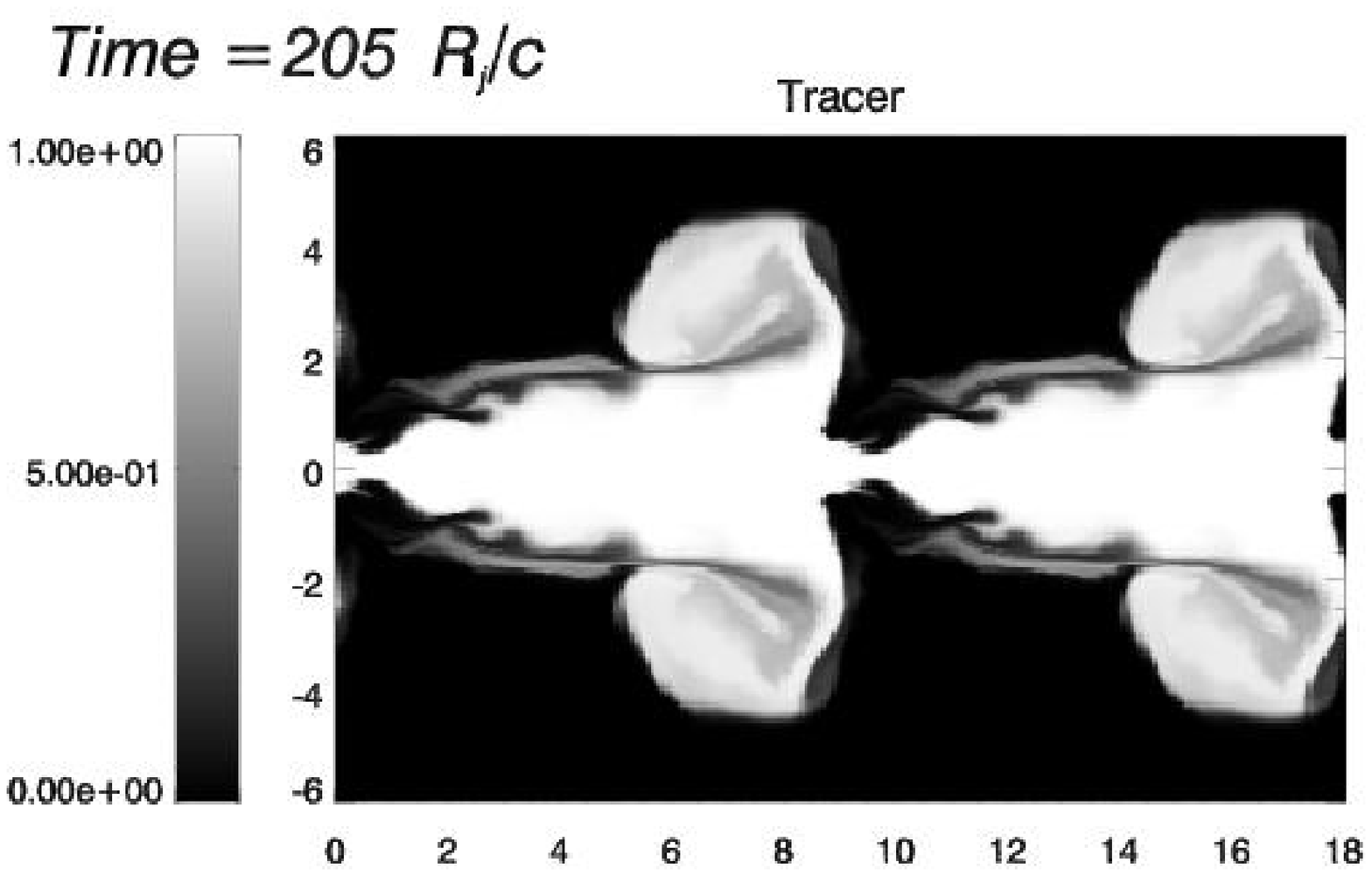,width=0.40
\textwidth,angle=0,clip=} \quad
\psfig{file=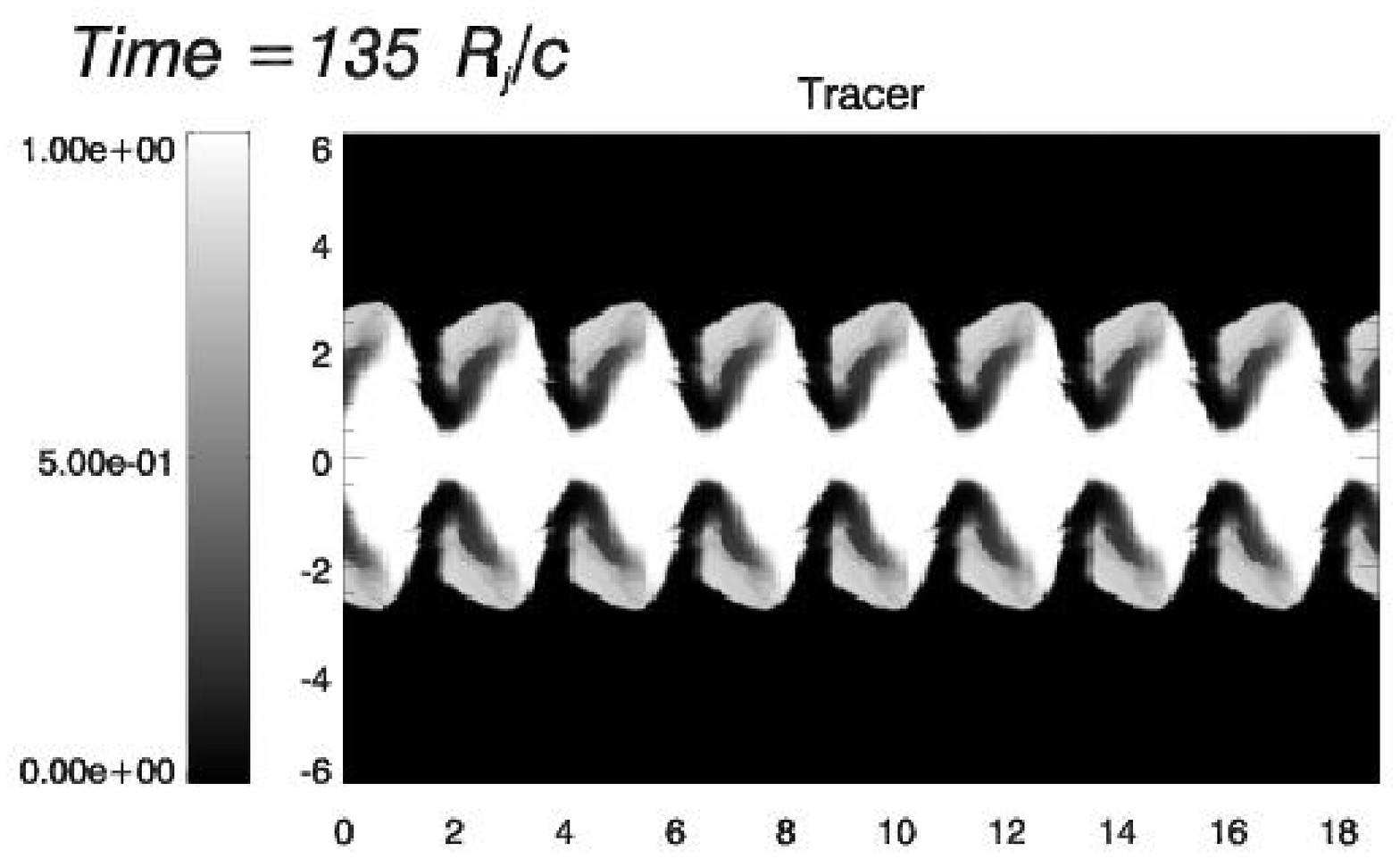,width=0.40 \textwidth,angle=0,clip=} }
\centerline{ \psfig{file=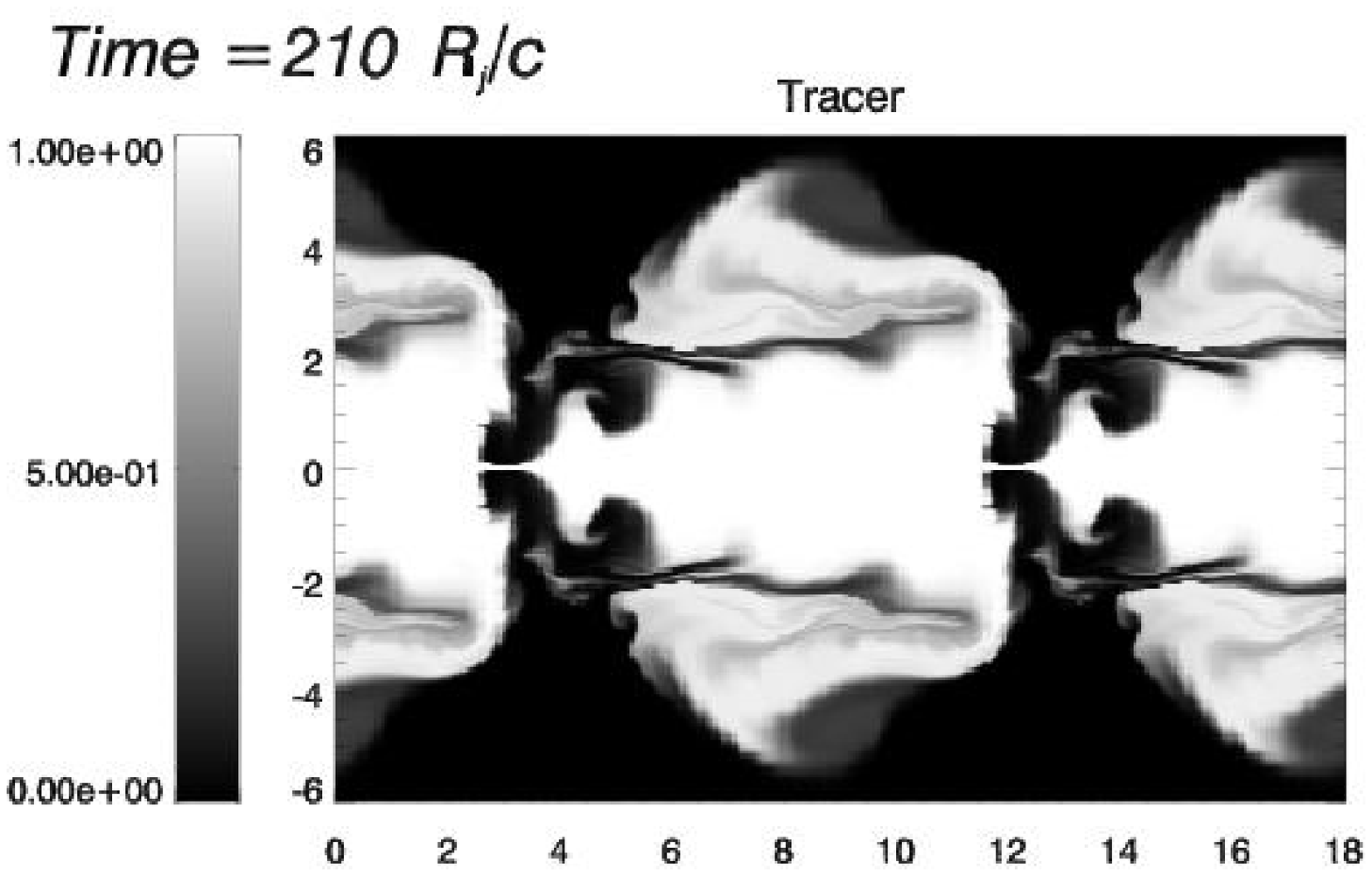,width=0.40
\textwidth,angle=0,clip=} \quad
\psfig{file=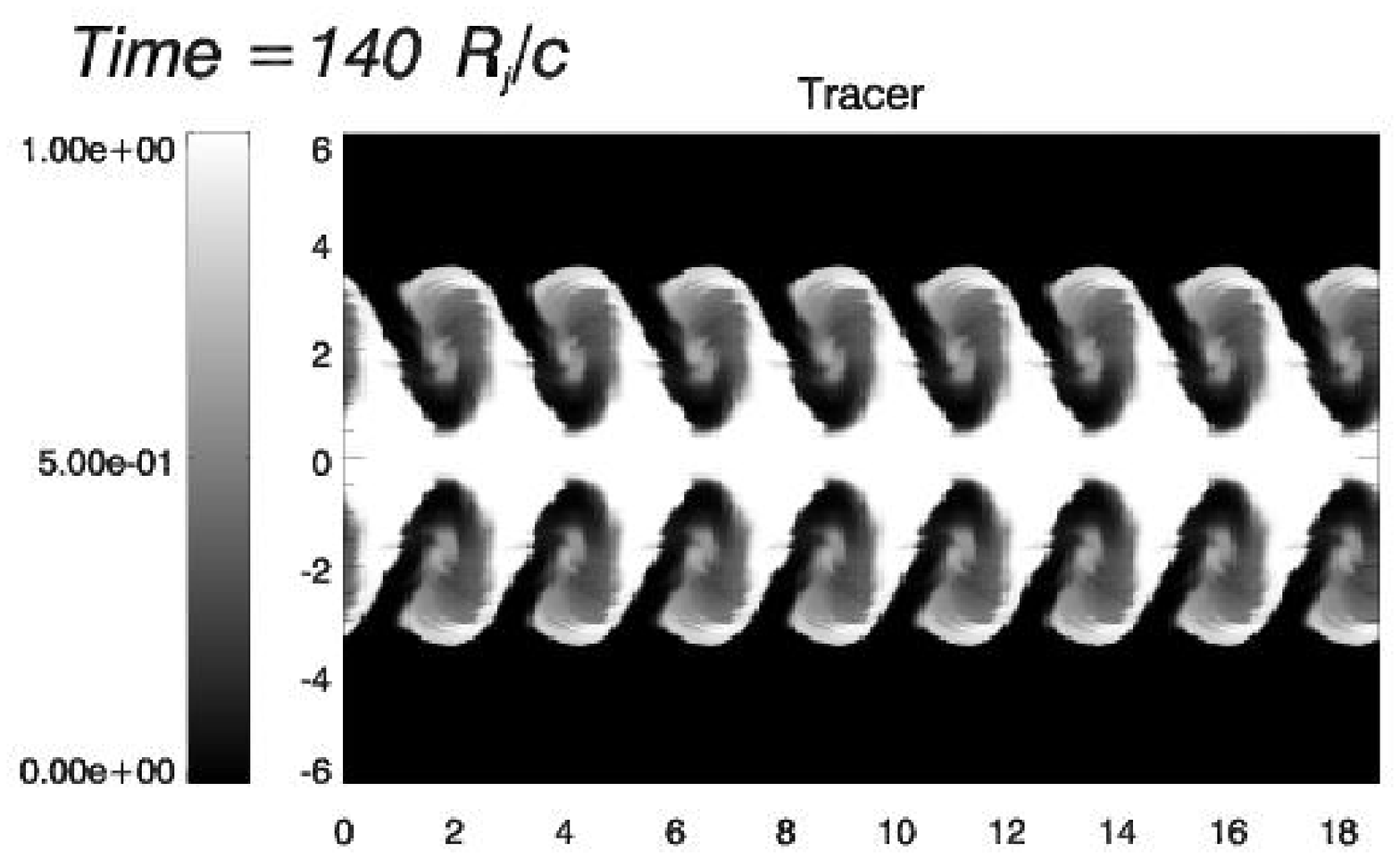,width=0.40 \textwidth,angle=0,clip=} }
\caption{Evolution of the jet particle fraction showing the
  development of mixing in two representative models. Left column
  (model B05): the ambient material carves its way through the jet
  difficulting the advance of the jet material which is suddenly
  stopped. Right column (model D05): the amount of ambient matter
  hampering the jet material is smaller and matter from the jet at the
  top of the jet crests is ablated by the ambient wind.  }
\label{fig:mix}
\end{figure*}

  The second mechanism of mixing starts during the transition to the
full non-linear regime and seems to act mainly in those models with
large $\Delta_{\rm peak}$. As we shall see below, this large value of
$\Delta_{\rm peak}$ is associated with the generation of a shock at
the jet/ambient interface at $t_{\rm peak}$, which appears to be the
responsible of the generation of wide mixing layers in those
models. Figure~\ref{fig:mix} shows a sequence of models with the
evolution of mixing in two characteristic cases, B05 and D05, during
the late lapse of the saturation phase. The evolution of model B05 is
representative of models A05, B05 and B10.  Models B20, C05, C10, C20
and D10 have evolutions closer to model D05. As it is seen in
Fig.~\ref{fig:mix}, in the case of model B05 (left column panels), the
ambient material carves its way through the jet difficulting the
advance of the jet material which is suddenly stopped. The result is
the break-up of the jet. In model D05, (right column panels), matter
from the jet at the top of the jet {\it crests} is ablated by the
ambient wind forming vortices of mixed material filling the valleys.

\begin{figure*}
\centerline{ \psfig{file=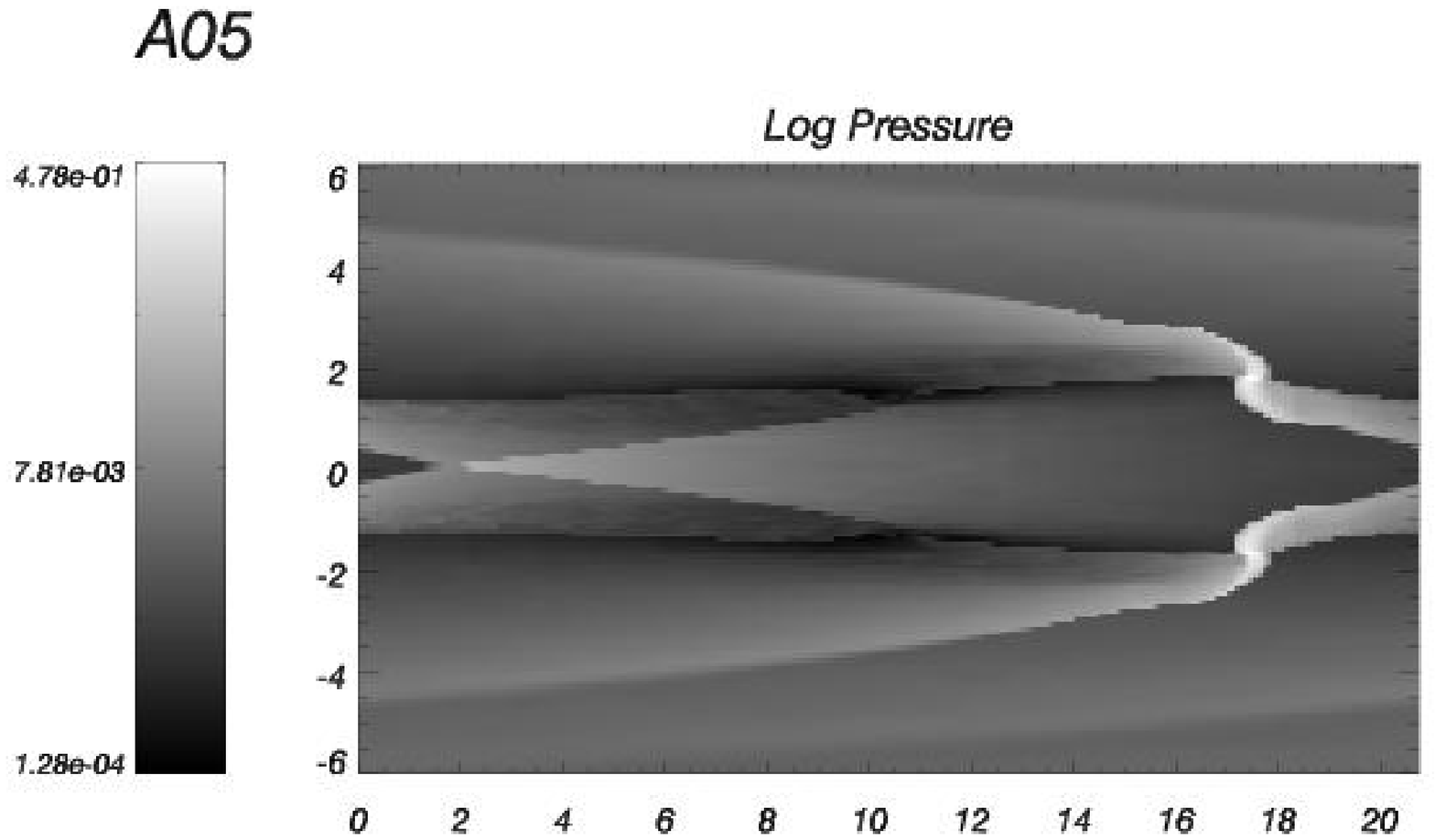,width=0.45
\textwidth,angle=0,clip=} \quad
\psfig{file=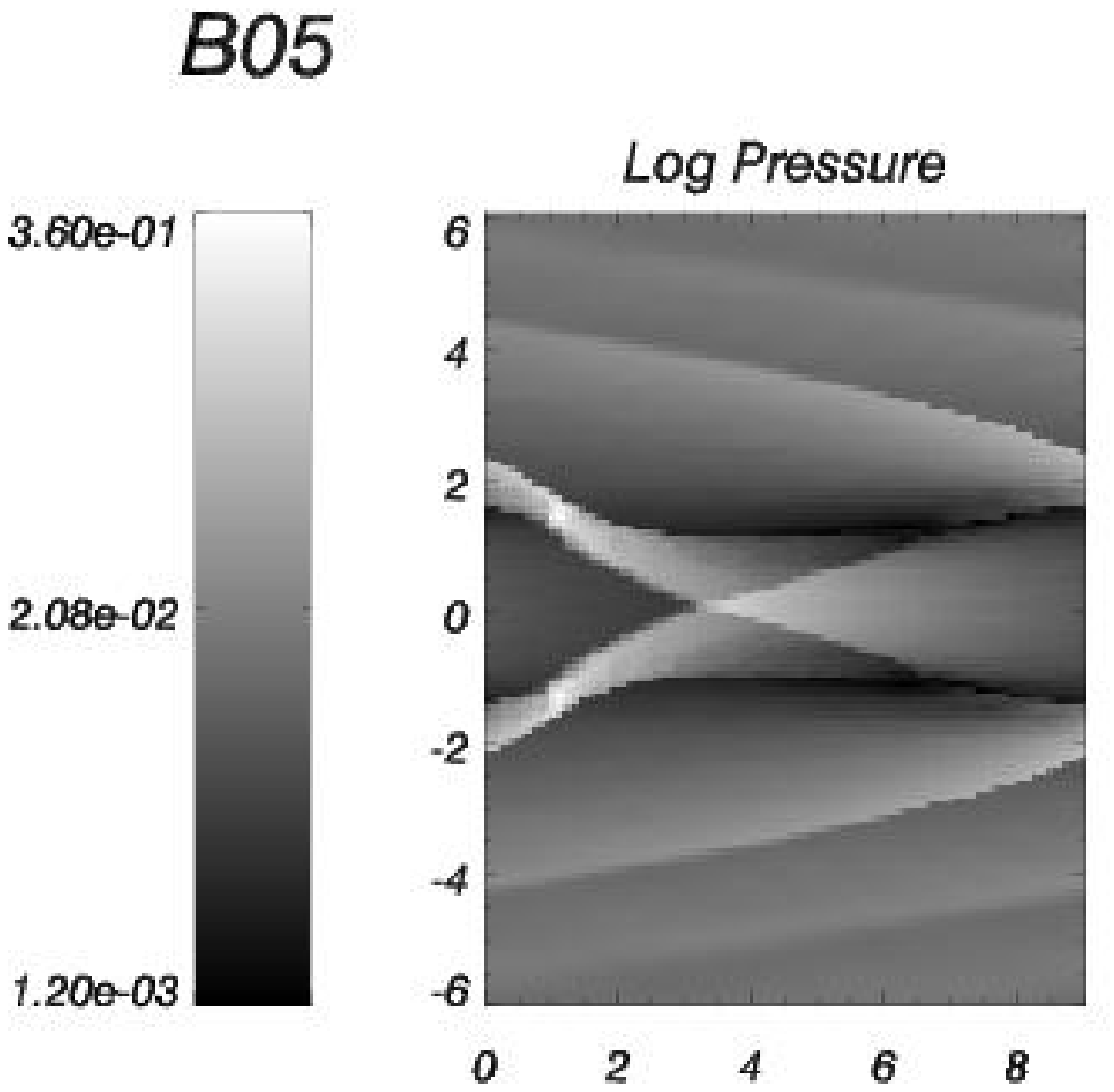,width=0.45 \textwidth,angle=0,clip=} }
\centerline{ \psfig{file=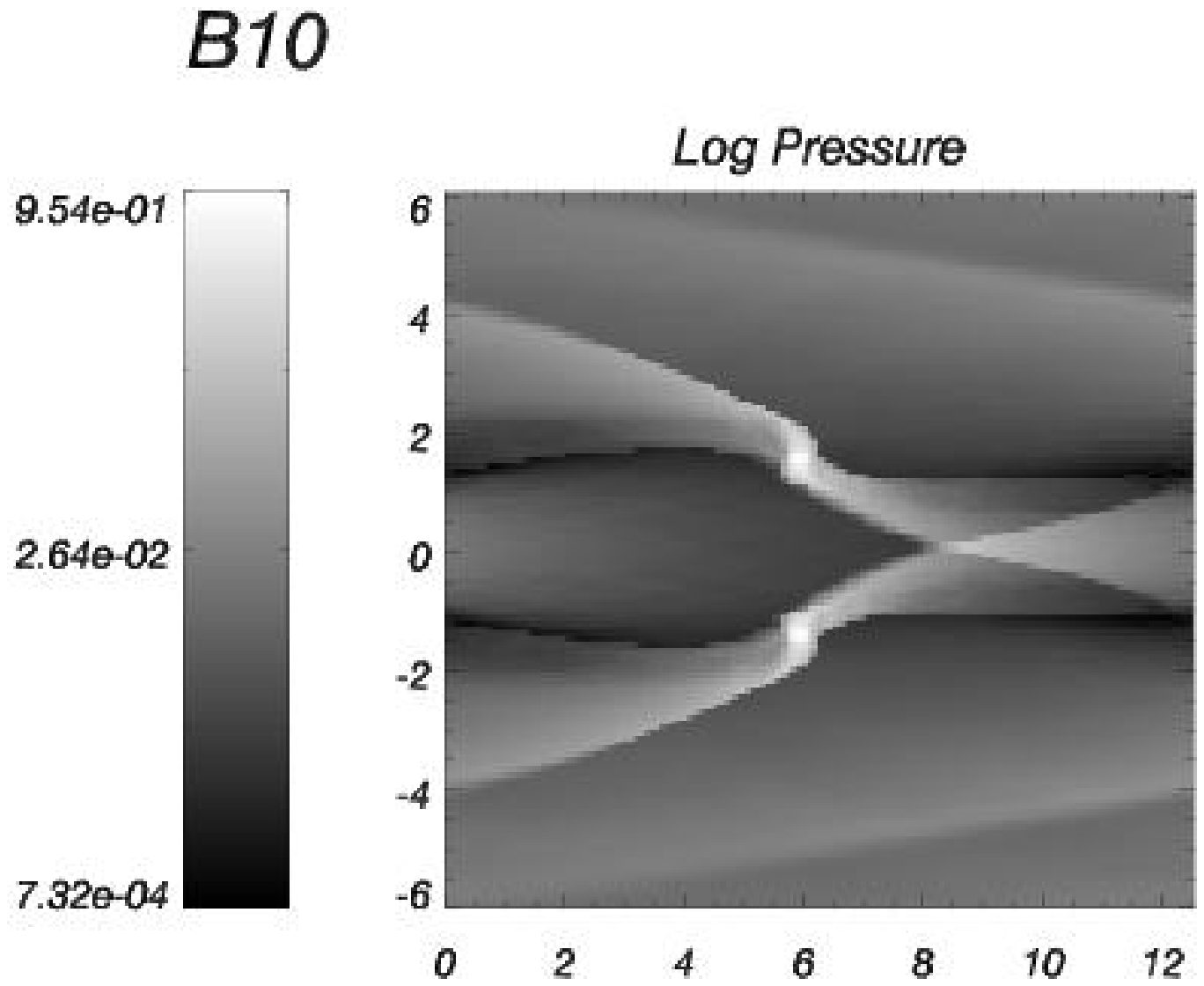,width=0.45
\textwidth,angle=0,clip=} \quad
\psfig{file=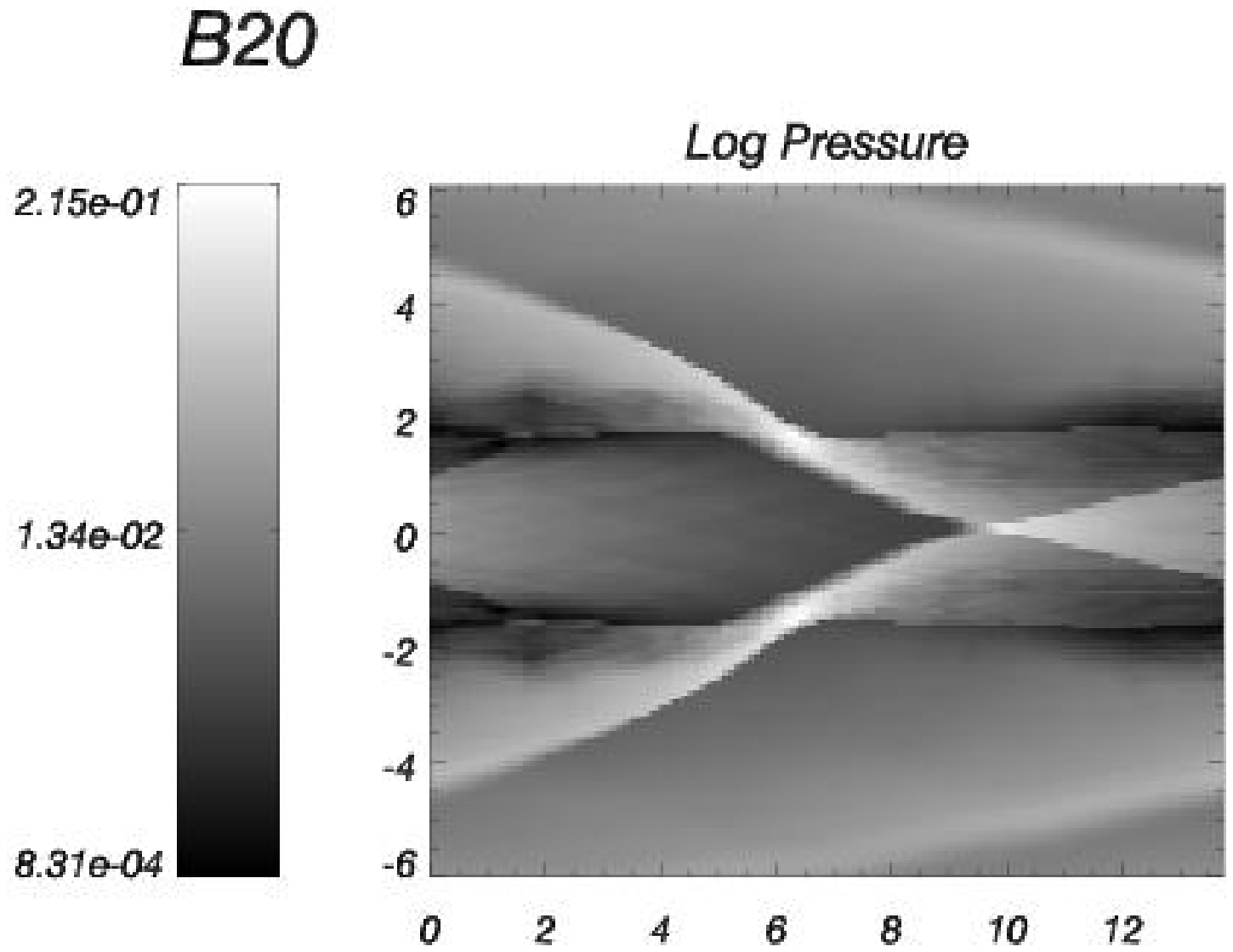,width=0.45 \textwidth,angle=0,clip=} }

\caption{Pressure distribution at the onset of the jet/ambient surface
  distortion at the end of the saturation phase for models A05, B05,
  B10 and B20. The corresponding times are 355, 190, 370 and 765
  $R_j/c$. In the case of models A05, B05 and B10 this distortion
  leads to the formation of a shock.  
}
       
\label{fig:preshock}
\end{figure*}

  The large amplitude of $\Delta_{\rm peak}$ reached in models A05,
B05 and B10 is clearly associated with a local effect occurring in the
jet/ambient interface (see second panel in the left column of
Fig.~\ref{fig:mix} and Fig.~\ref{fig:preshock}) that leads to the jet
disruption.  The sequence of events preceding the jet disruption
includes the formation of oblique shocks at the end of the
linear/saturation phase, the local effect in the jet/ambient
interface just mentioned and then a supersonic transversal expansion
of the jet that leads to i) a planar contact discontinuity (see last
panels in the left column of Fig.~\ref{fig:mix}), and ii) the
formation of a shock (see below) that propagates traversally (see
Sect.~\cite{ss:transversal}). It appears that contrary to the velocity
perturbations in the jet reference frame (see Fig.~1 of Paper I), the
maximum relative amplitudes of pressure perturbation are strongly
dependent on physical parameters of simulations\footnote{In order to
see how much this peak in relative pressure amplitude was influenced
by initial relative amplitude to background values, we repeated the
simulation corresponding to model D05 with the same initial relative
amplitude as model A05 (i.e., three orders of magnitude larger).
Results show that there is not a significant difference in the peak
values of the pressure and that the evolution is basically the same as
the one of the original model.}.  The origin of the shock in models
with $\Delta_{\rm peak} \approx 100$, that enhances the turbulent
mixing of the jet and ambient fluids, can be found in the nonlinear
evolution of Kelvin-Helmholtz instability that leads to significant
changes of local flow parameters at the end of saturation phase. For
instance, the oblique shock front resulting from the steepening of
sound waves, during the linear and saturation phases (see Figs.~3-6 of
Paper I), crosses the initial shear layer at the interface of jet and
ambient medium.  The formation of such oblique shock implies a sudden
and local growth of gradients of all the dynamical quantities at the
jet boundary. The local conditions changed by these oblique shocks may
become favourable for the development of other instabilities like
those discussed by Urpin (2002).

\begin{figure*} 
\centerline{
\psfig{file=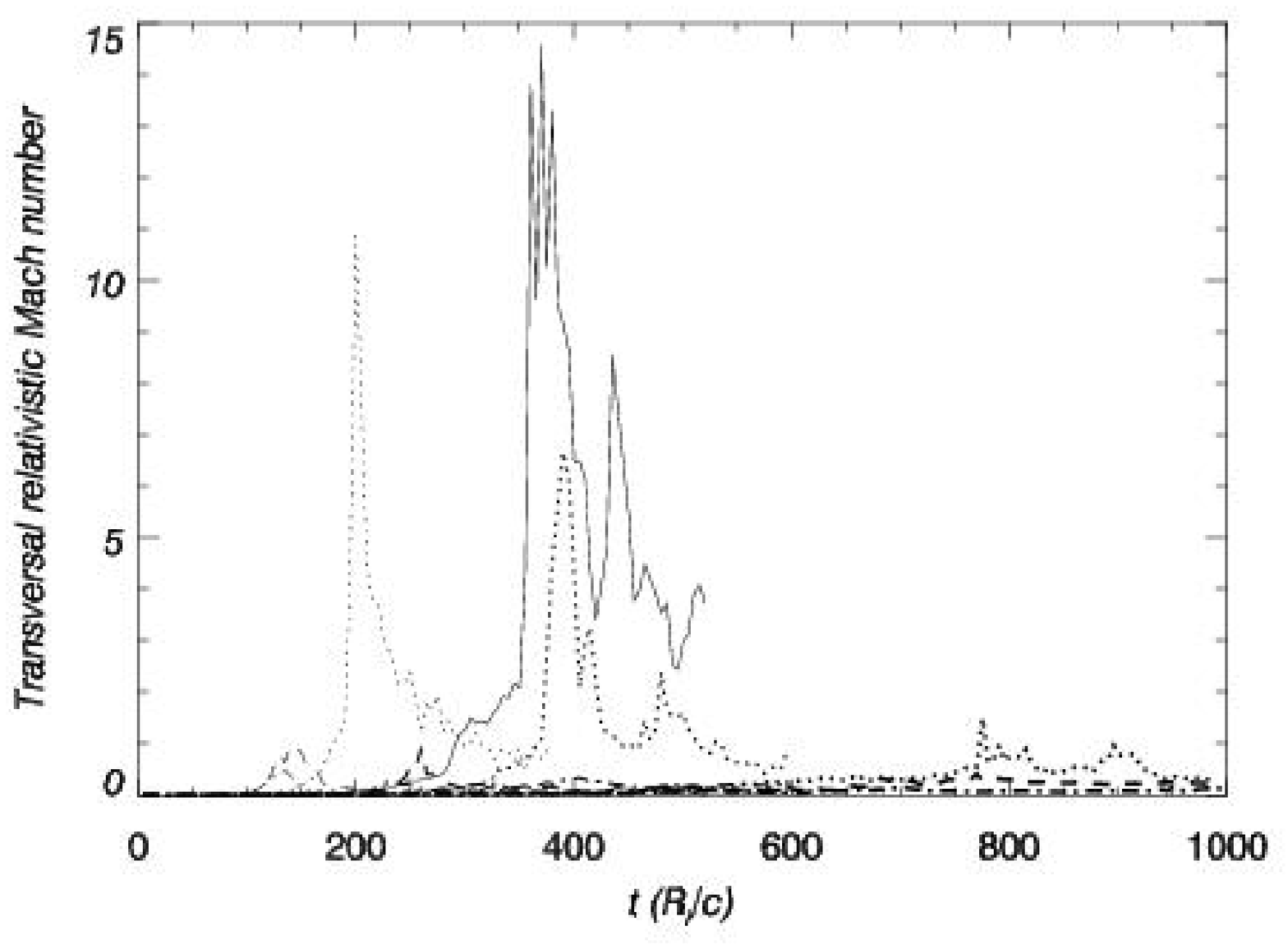,width=0.8\textwidth,angle=0,clip=0} 
}

\caption{Time evolution of the maxima of the transversal Mach number
  of the jet with respect to the unperturbed ambient medium, {\cal
  M}$_{j,\perp}$. See text for further explanations. Lines are as in
  Figure \ref{fig:tracerdisp}.  }

\label{fig:fmach}
\end{figure*}

  The generation of the shock wave at the jet/ambient interface
is reflected in the evolution of the maxima of {\cal M}$_{j,\perp} =
\gamma_{j,\perp}v_{j,\perp}/(\gamma_{c_{sa}} c_{sa})$
($\gamma_{j,\perp}$ and $\gamma_{c_{sa}}$ being the Lorentz factors
associated with $v_{j,\perp}$ and $c_{sa}$, respectively) representing
the transversal Mach number of the jet with respect to the
unperturbed ambient medium. This quantity becomes larger than 1 around
$t_{\rm peak}$ in those models with $\Delta_{\rm peak} \approx 100$
(see Fig.~\ref{fig:fmach}) pointing toward a supersonic expansion of
these jets at the end of the saturation phase. The fact that in all
our simulations the ambient medium surrounding colder models (i.e., A,
B) are also colder (see Sect.~\ref{sect:numsim}) favors the generation
of shocks in the jet/ambient interface in these models. On the other
hand, in the case of models with the highest Lorentz factors (B20,
C20, D20) the transversal velocity can not grow far enough to generate
the strong shock which breaks the slower jets.

  In order to extend our conclusions to a wider region in the initial
parameter space, we have performed supplementary simulations (namely
F, G, H, I, J, K, L) with the aim of clarifying the effect of the
ambient medium in the development of the disruptive shock appearing
after saturation.  In all the cases, the external medium is that of
model A05. 

  From models F to H, internal energy in the jet is increased and
rest-mass density decreased in order to keep pressure equilibrium,
whereas the jet Lorentz factor is kept constant to its value in model
A05. The transversal Mach numbers at the peak in these models reach
values very similar to that of model A05 ($\simeq 14$). In fact, the
formation of a shock at the end of the saturation phase is observed in
these models as it is in model A05.

  The initial momentum density in the jet has decreased along the
sequence A05, F, G, H. Simulations I and J have the same
thermodynamical values as models F and G, respectively, but have
increasing jet Lorentz factors to keep the same initial momentum
density as model A05. In the case of model I we find the same behavior
as in previous ones: large transversal Mach number, shock and
disruption. On the contrary, model J behaves much more like model B20,
with a value of transversal Mach number slightly larger than one,
strong expansion and almost no mixing. Finally, in order to know to
which extent this change in behavior was caused by the increase in
Lorentz factor or in specific internal energy, in simulations K and L
we cross these values with respect to those in I and J. The results
from these last simulations show that the evolution of model M is very
close to that of model I: the large value of the transversal Mach number 
at the peak, shock and strong mixing; and that the evolution of
model K is close to that of model J: a weaker shock, expansion and no
mixing. 

  In the models developing a shock, mixing is associated with vorticity
generated after the shock formation and, in the case of models A05,
B05, B10 as well as F, G, H, I and L matter from the ambient
penetrates deep into the jet as to reach the jet axis (jet
break-up). The times at which this happens for the different models
are displayed in Table~\ref{tab:phases}. Note that in models A05 and
B05 the entrainment of ambient matter up to the axis occurs just after
$t_{\rm peak}$, whereas in model B10 occurs later, probably because
the shock in this model is weaker. The process of mixing can be
affected by resolution, as small resolutions may suppress the
development of turbulence. We have analyzed this in the Appendix in
which we focus on the influence of longitudinal resolution.

\subsection{Fully nonlinear evolution: jet/ambient momentum transfer
\label{ss:mtransfer}}

\begin{figure*} 
\centerline{
\psfig{file=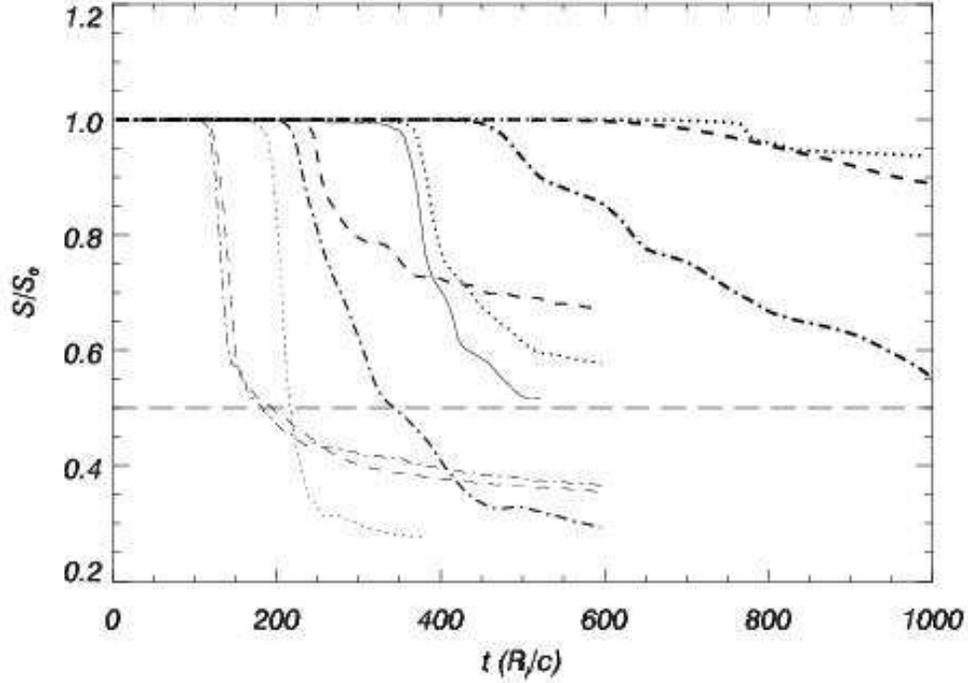,width=0.8\textwidth,angle=0,clip=0} 
}

\caption{Evolution of the total longitudinal momentum in the jet as a
  function of time for all the simulations. Lines represent the same
  models than in Fig.~\ref{fig:tracerdisp}. The long-dashed horizontal
  line serves us to identify those models tranferring more than 50\%
  of the initial jet momentum to the ambient.}
       
\label{fig:axialm}
\end{figure*}

  Let us now analyze the evolution of the longitudinal momentum in the
jets as a function of time. Figure~\ref{fig:axialm} shows the
evolution of the total longitudinal momentum in the jet for the
different models. Jets in models B05, C05, D05 and D10 (also D20)
transfer more than 50\% of their initial longitudinal momentum to the
ambient, whereas models A05, B10 and C10 (also B20) seem to have
stopped the process of momentum transfer retaining higher fractions of
their respective initial momenta. Models C20 and D20 continue the
process of momentum exchange at the end of the simulations but at a
remarkable slower rate (specially C20).

\begin{figure*} 
\centerline{
\psfig{file=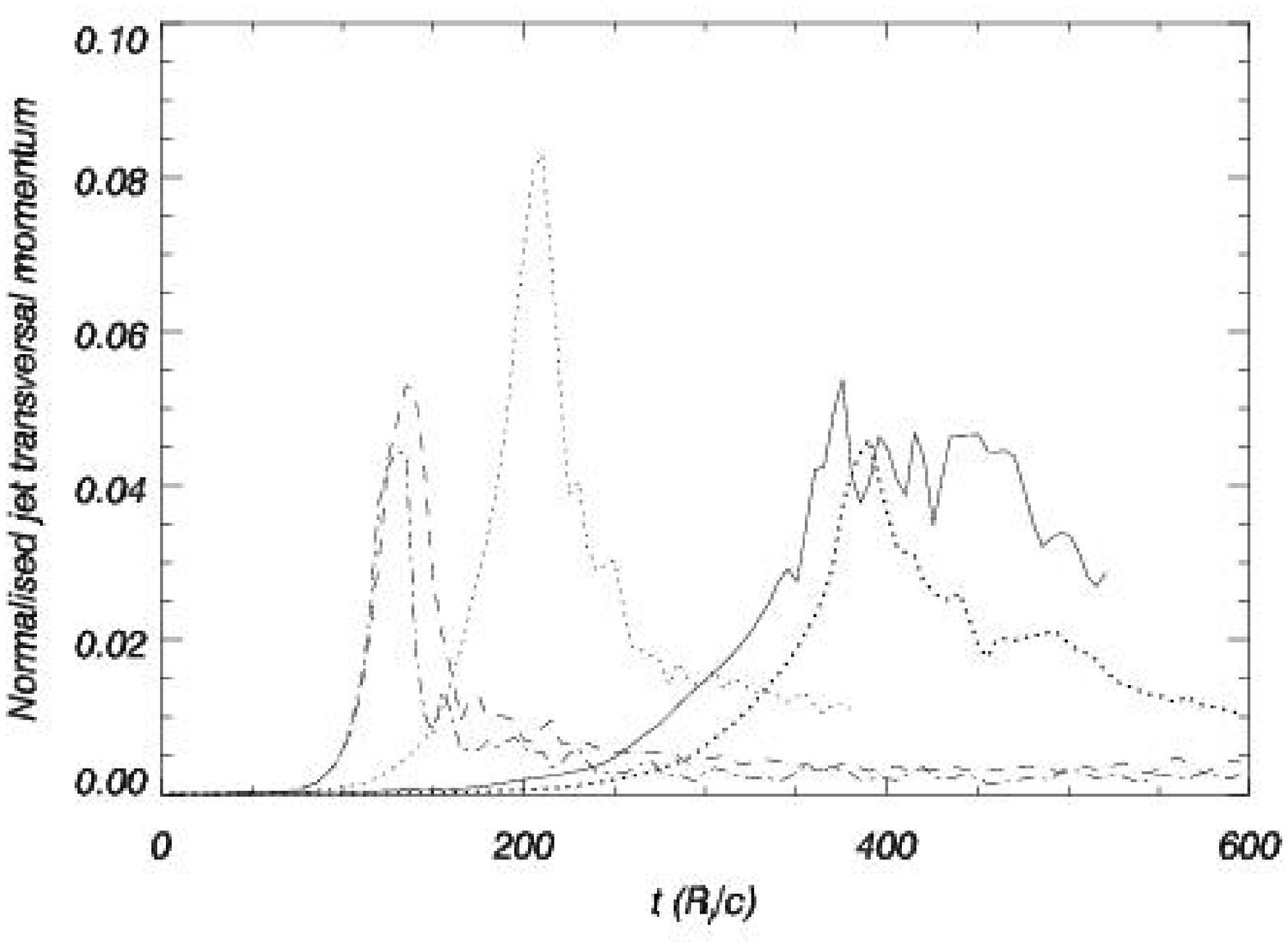,width=0.45 \textwidth,angle=0,clip=} \quad
\psfig{file=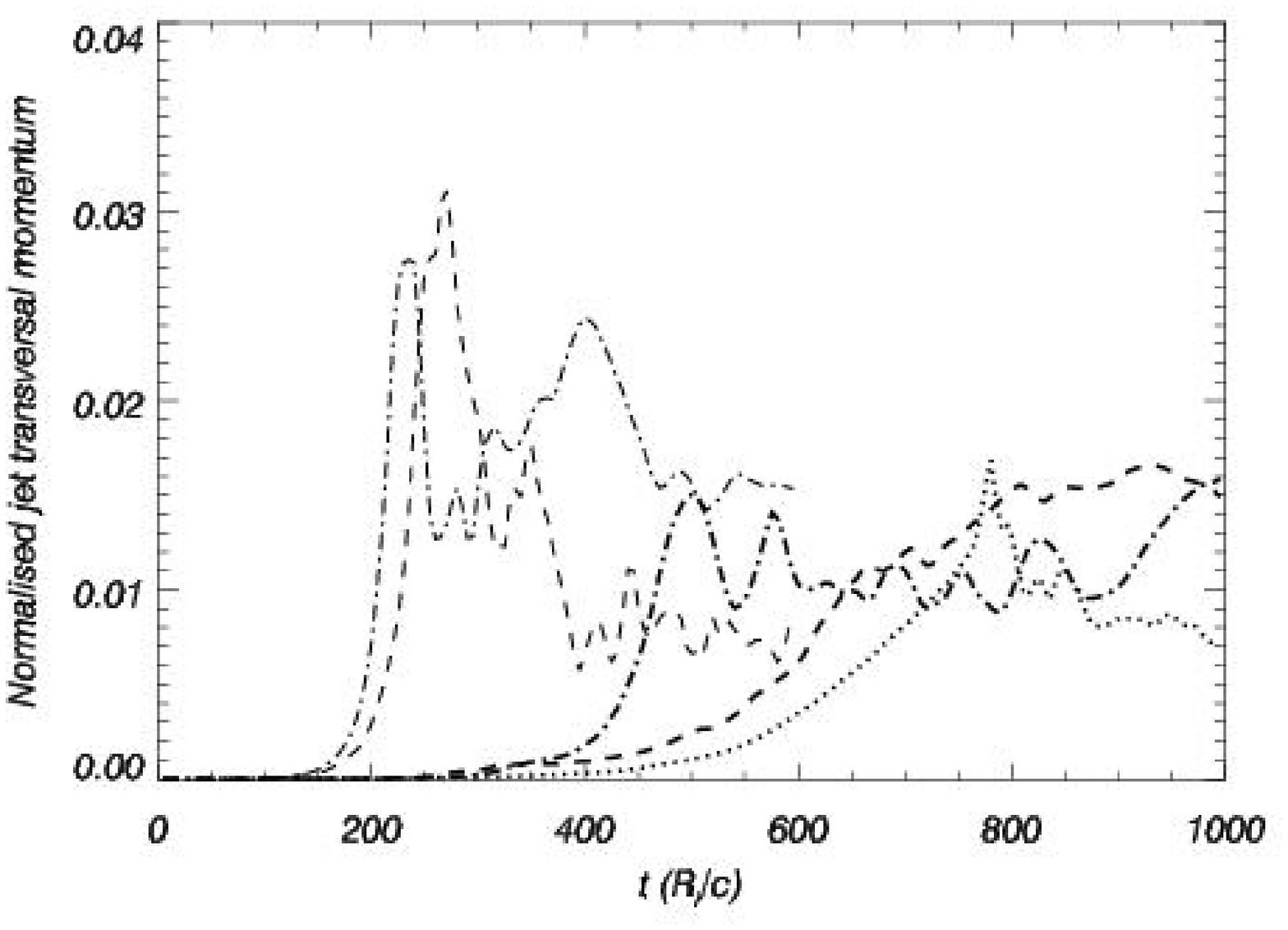,width=0.45 \textwidth,angle=0,clip=} \quad
}
\caption{Evolution of the total transversal momentum in the jet as a
  function of time for all the simulations. Lines represent the same
  models than in Fig.~\ref{fig:tracerdisp}. Left panel: models A05,
  B05, B10, C05, D05. Right panel: B20, C10, C20, D10, D20. Note the
  change in both horizontal and vertical scales between the two
  panels.}
\label{fig:tmom}
\end{figure*}

  In the case of models F, G, H, I and L, the transfer of longitudinal
momentum is also very efficient. The reason why these models, as well
as A05, B05 and B10, develop wide shear layers and transfer more than
50\% of their initial momentum to the ambient could be turbulent
mixing triggered by the shock. In the case of models D10 and D20, the
processes of mixing and transfer of longitudinal momentum proceed at a
slower rate pointing to another mechanism. The plots of the time
evolution of the jet's transversal momentum (Fig.~\ref{fig:tmom}) for
the different models give us the answer.  Jets disrupted by the shock
(as A05, B05, B10) have large relative values of transversal momentum
at saturation ($>0.04$) that decay very fast afterward (A05 is an
exception).  The peak in transversal momentum coincides with the
shock formation and the fast lateral expansion of the jet at $t_{\rm
peak}$. Contrarily, models D10 and D20 have a sustained value of
transversal momentum after saturation which could drive the process of
mixing and the transport of longitudinal momentum.  In these models,
the originally high internal energy in the jet and the high jet
Lorentz factor (that allows for a steady conversion of jet kinetic
energy into internal) make possible the sustained values of
transversal momentum.  Between these two kinds of behavior are hot,
slow models C05, D05 that do not develop a shock having, then, thin
mixing layers, but transferring more than 50\% of their longitudinal
momentum.

\subsection{Fully nonlinear evolution: classification of the models
\label{ss:mtransfer}}

  Our previous analysis based in the width of the mixing layers and
the fraction of longitudinal momentum transfered to the ambient can be
used to classify our models:

\begin{itemize}

\item Class I (A05, B05, B10, F, G, H, I, L): develop wide shear
      layers and break up as the result of turbulent mixing driven by
      a shock.
\item Class II (D10, D20): develop wide shear layers and transfer more
      than 50\% of the longitudinal momentum to the ambient, as a
      result of the sustained transversal momentum in the jet after
      saturation.
\item Class III (C05, D05): have properties intermediate to models in
      classes I and II.
\item Class IV (B20, C10, C20, J, K): are the most stable.

\end{itemize}

\begin{figure*} 
\centerline{
\psfig{file=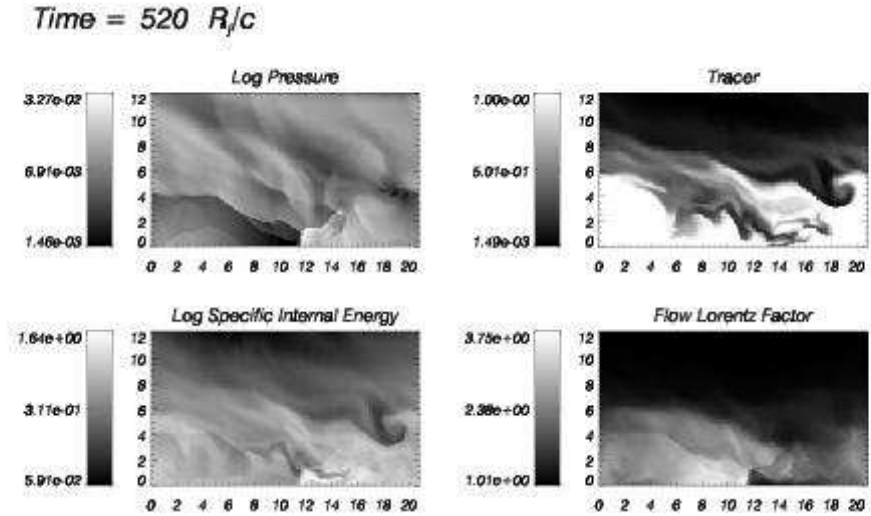,width=0.8\textwidth,angle=0,clip=0} 
}
\caption{Snapshot in the mixing phase of logarithmic maps of pressure, jet mass fraction and specific internal energy and non-logarithmic Lorentz factor for model A05. Only the top half of the jet is shown.}
\label{fig:fab05m}

\end{figure*}

\begin{figure*}
\centerline{
\psfig{file=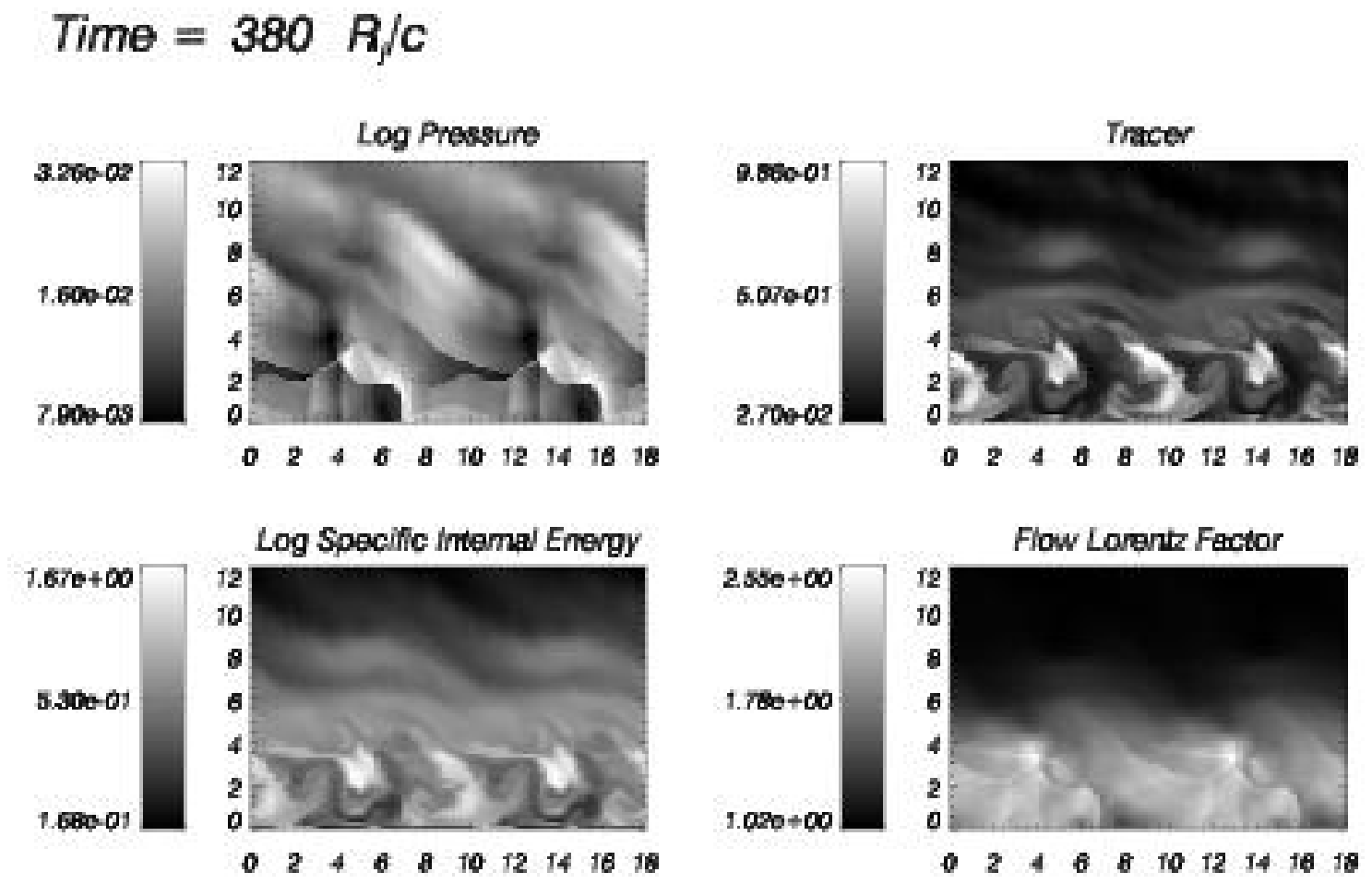,width=0.8\textwidth,angle=0,clip=0} 
}
\centerline{
\psfig{file=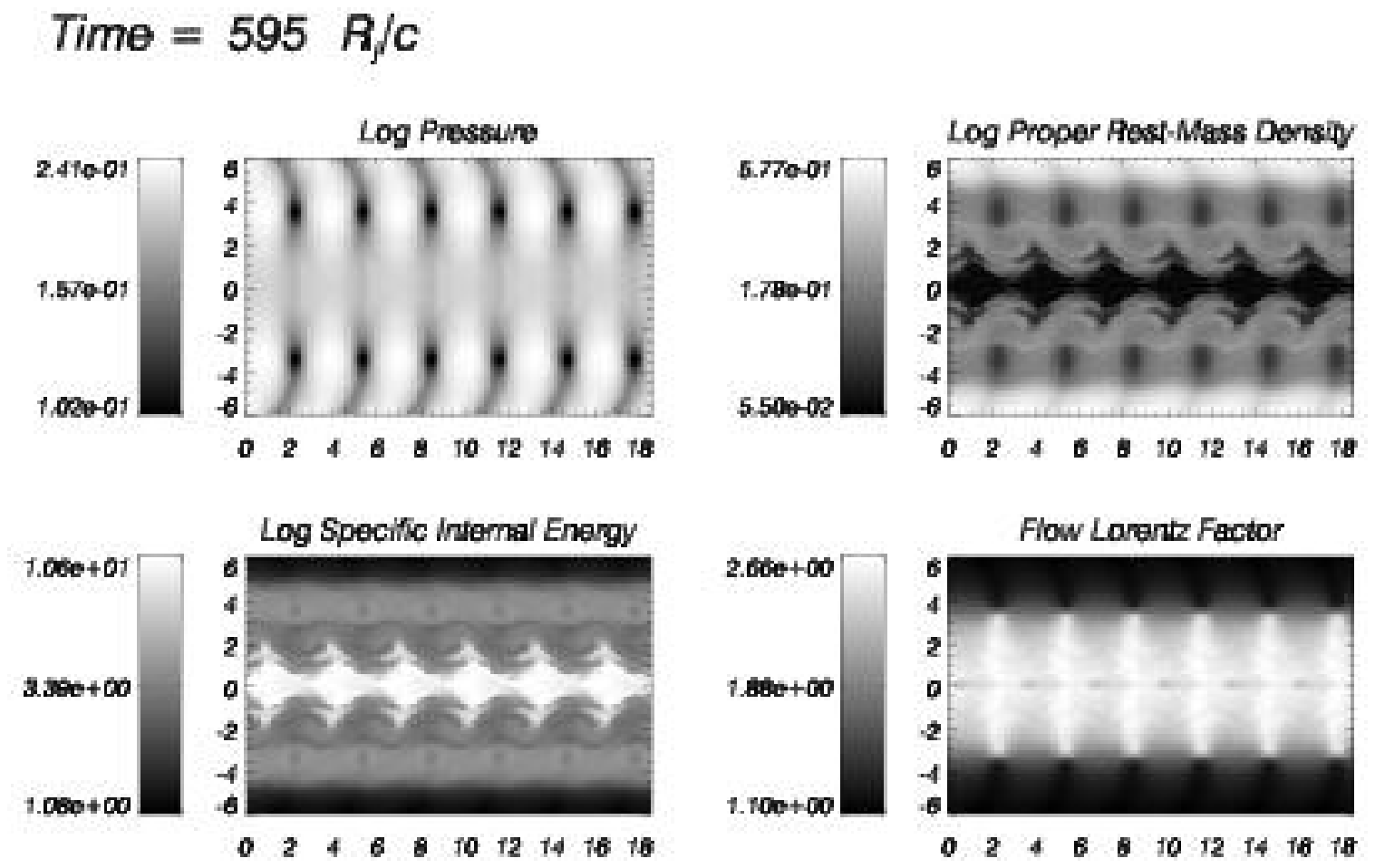,width=0.8\textwidth,angle=0,clip=0} 
}
\centerline{
\psfig{file=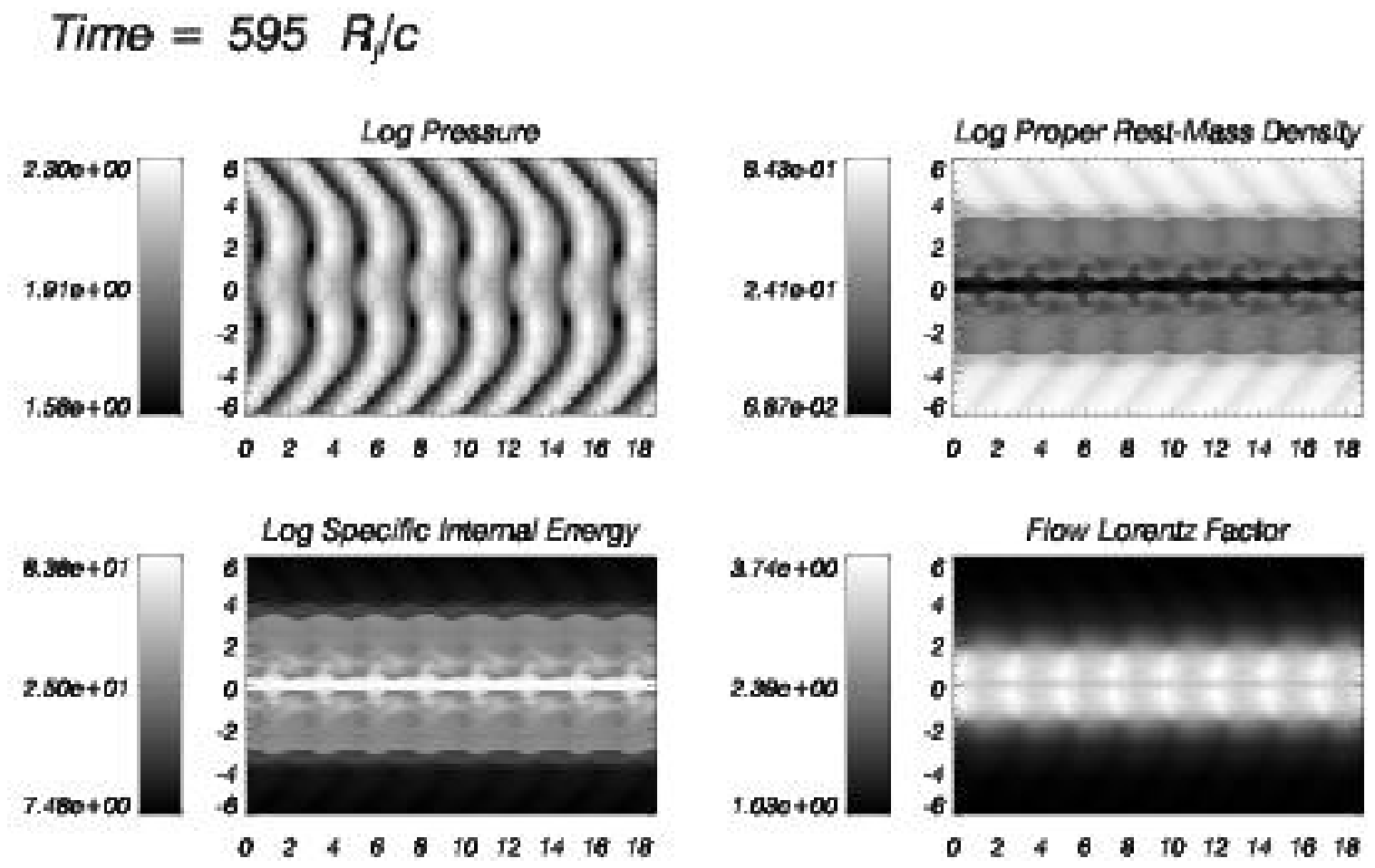,width=0.8\textwidth,angle=0,clip=0} 
}
\caption{Same as Fig. (\ref{fig:fab05m}) for models B05 (upper; only
  top half of the model shown), C05 (middle) and D05 (lower).}
\label{fig:fcd05m}

\end{figure*}

\begin{figure*}
\centerline{
\psfig{file=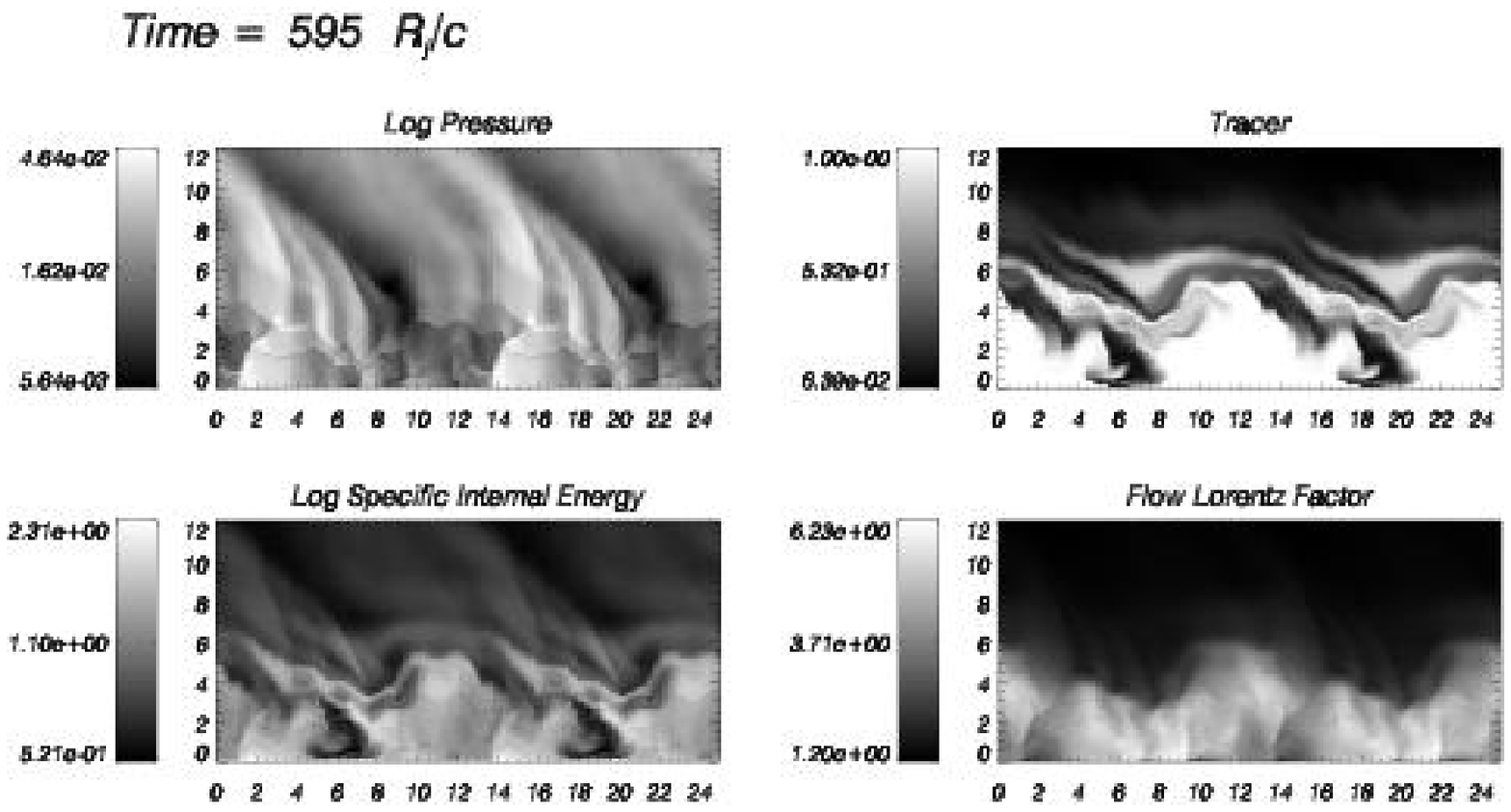,width=0.8\textwidth,angle=0,clip=0} 
}
\centerline{
\psfig{file=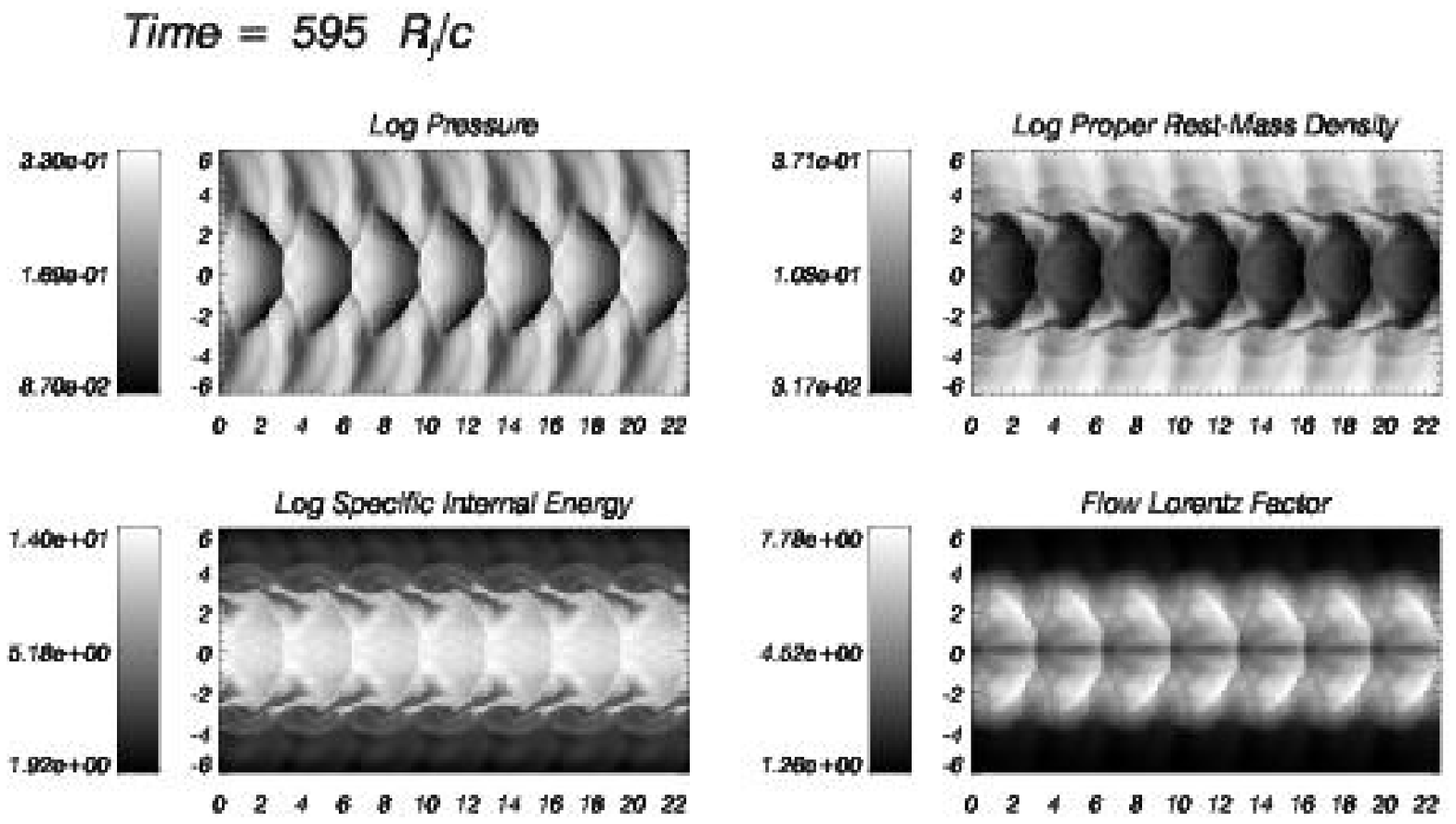,width=0.8\textwidth,angle=0,clip=0} 
}
\centerline{
\psfig{file=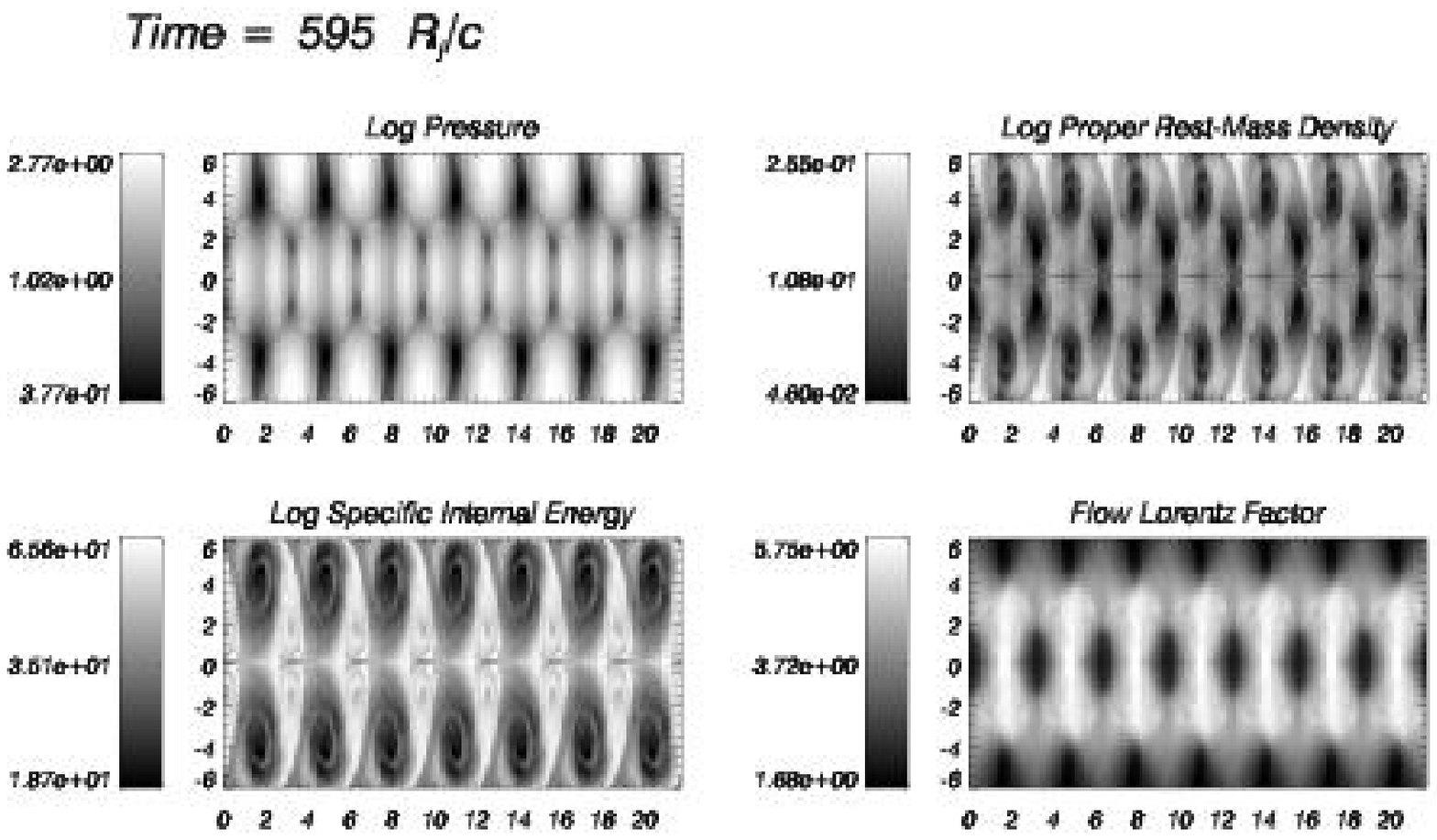,width=0.8\textwidth,angle=0,clip=0} 
}
\caption{Same as Fig. (\ref{fig:fab05m}) for models B10 (upper; only
  top half of the model shown), C10 (middle) and D10 (lower; only top
  half of the model shown).}

\label{fig:fbc10m}

\end{figure*}

\begin{figure*}
\centerline{
\psfig{file=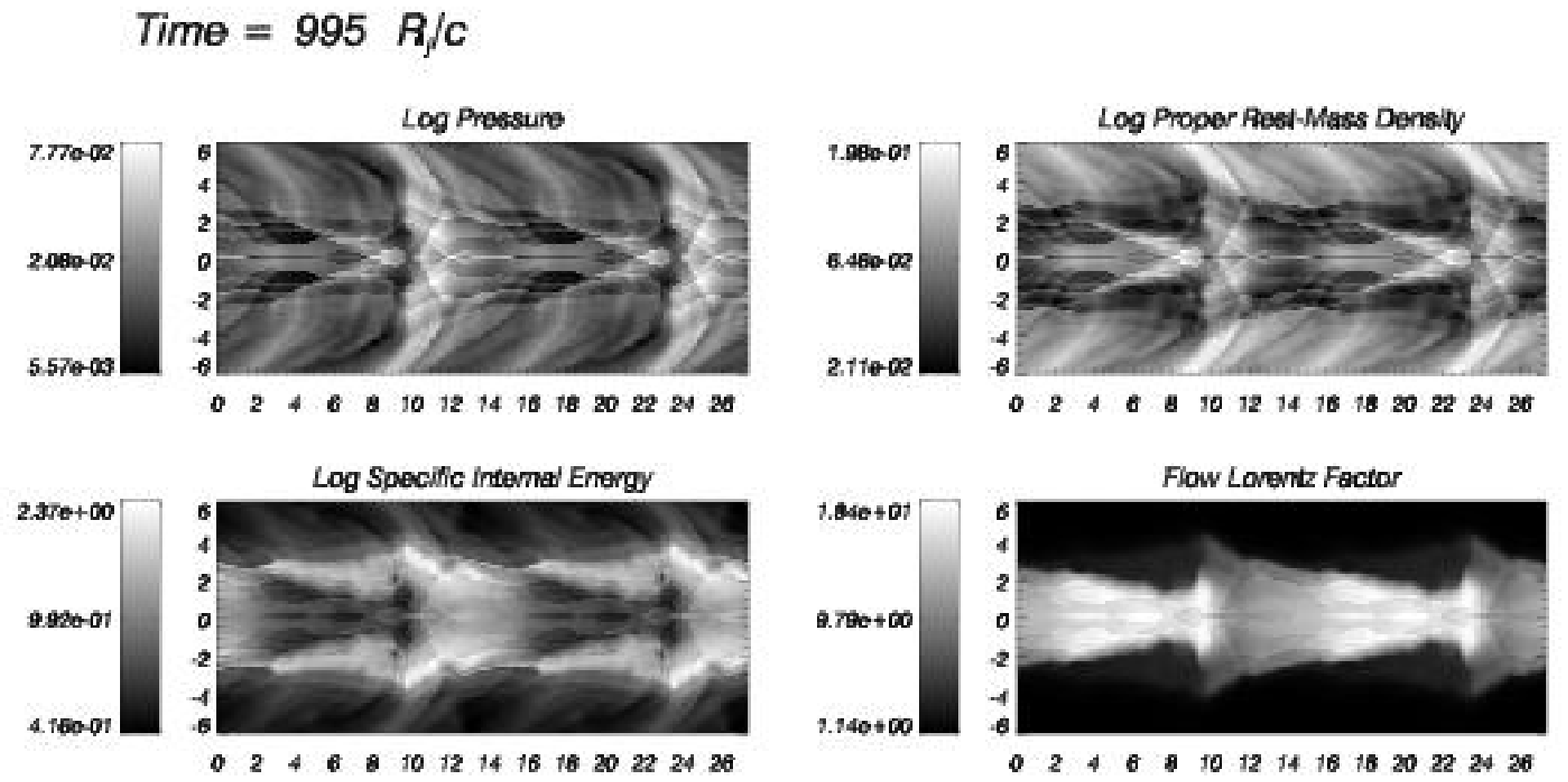,width=0.8\textwidth,angle=0,clip=0} 
}
\centerline{
\psfig{file=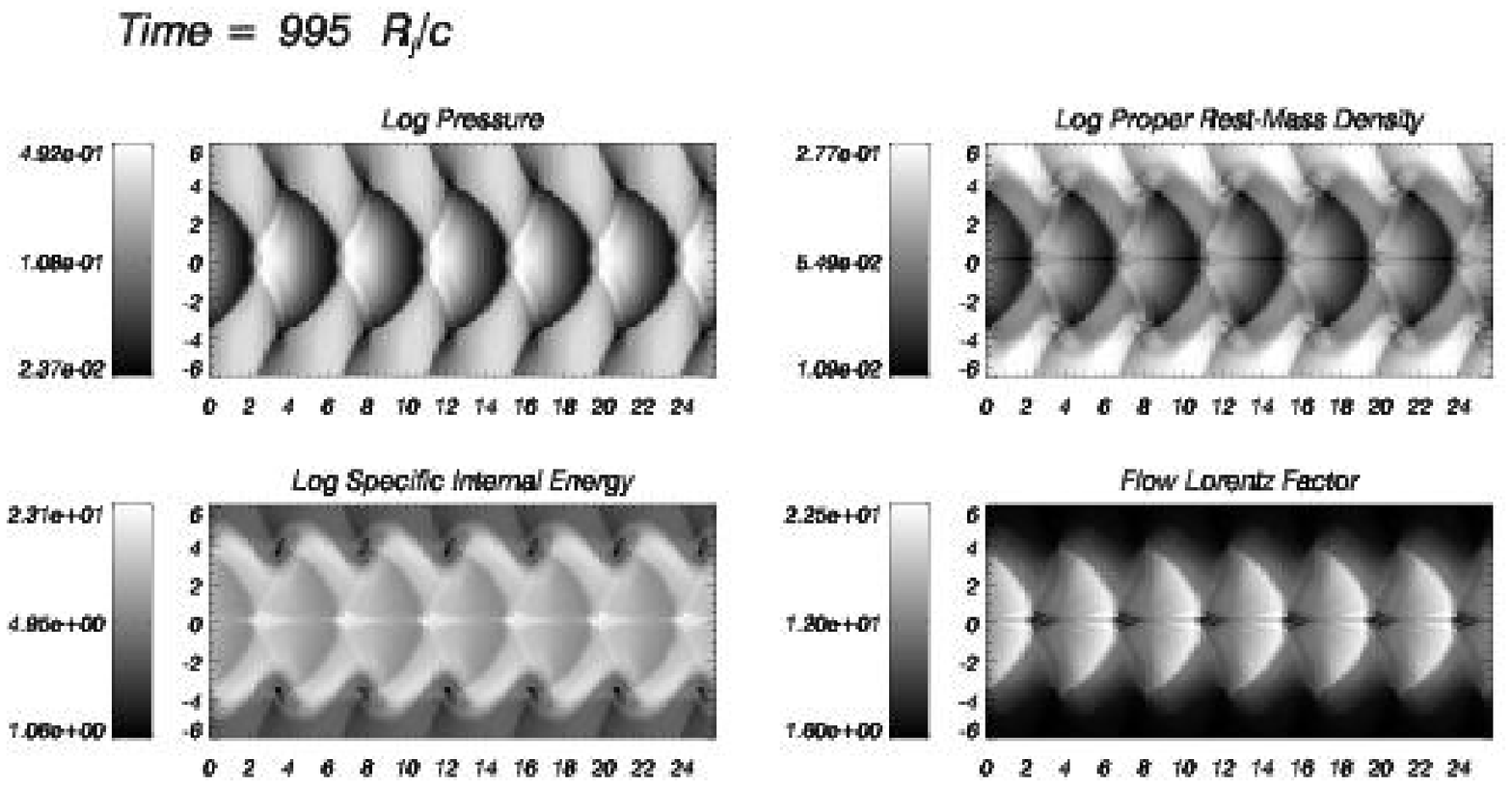,width=0.8\textwidth,angle=0,clip=0} 
}
\centerline{
\psfig{file=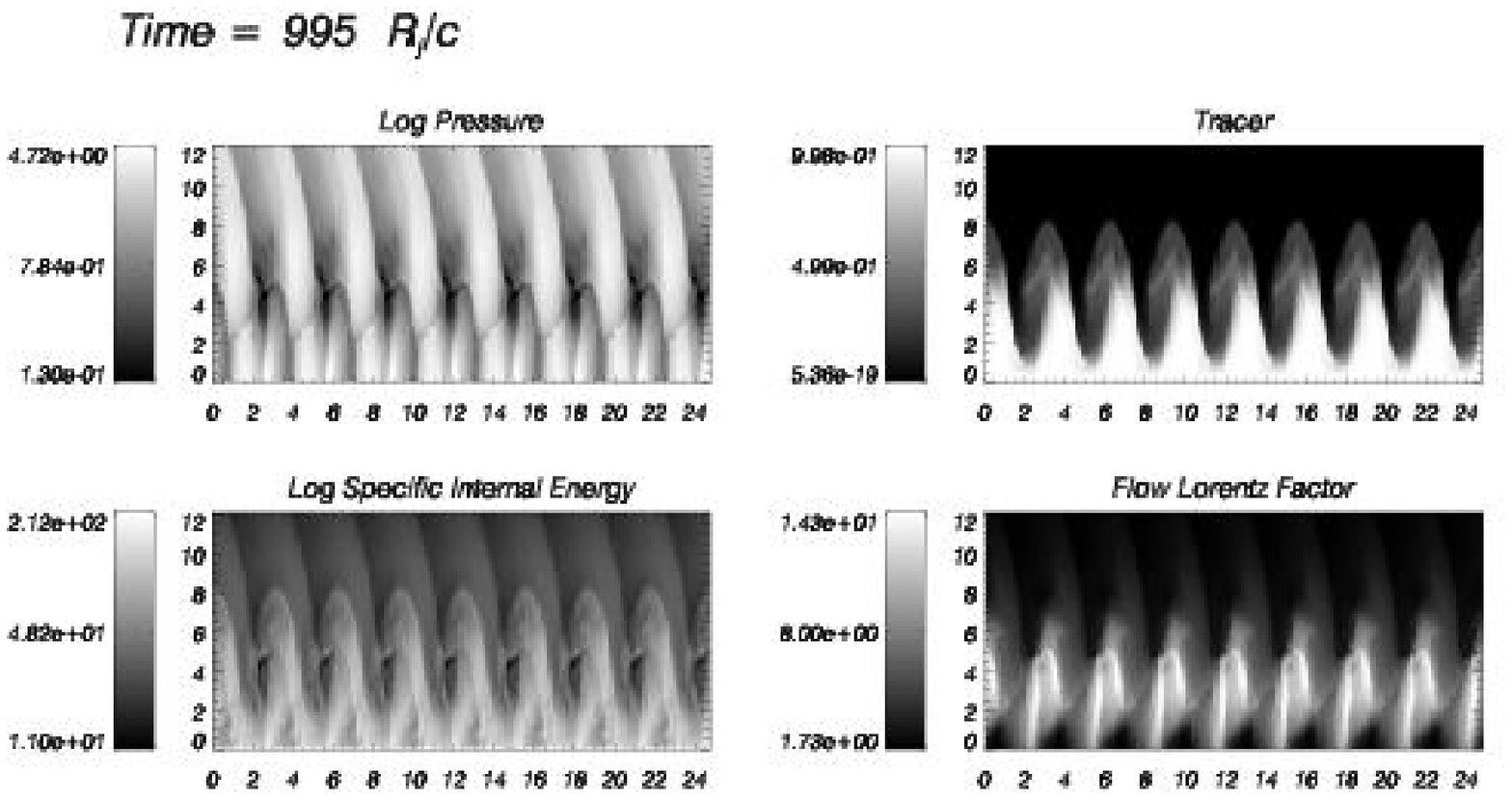,width=0.8\textwidth,angle=0,clip=0}
}
\caption{Same as Fig. (\ref{fig:fab05m}) for models B20 (upper), 
C20 (middle) and D20 (lower; only
  top half of the model shown).} 
\label{fig:fcd20m}

\end{figure*}

  Figures~\ref{fig:fab05m}-\ref{fig:fcd20m} show the flow structure of
the different models at the end of the simulations. The following
morphological properties of the members of each class are remarkable:

\begin{itemize}

\item Class I: irregular turbulent pattern of the flow, the structure
of KH modes still visible on the background of the highly evolved mean
flow pattern.

\item Class II: a regular pattern of "young" vortices (visible in the
tracer and specific internal energy distributions), the structure of
KH modes visible. The enhanced transfer of momentum found in the
models of this class is probably connected to the presence of these
"young" vortices.

\item Class III: the flow is well mixed, i.e. tracer, internal energy
and Lorentz factor are smoothed along lines parallel to the jet
symmetry plane. Highly evolved vortices visible. A fossil of KH modes
visible only as pressure waves.

\item Class IV: no vortices, no chaotic turbulence, very weak mixing,
very regular structure of KH modes.

\end{itemize}

\subsection{Transversal jet structure at late stages of evolution}
\label{ss:transversal}

  At the end of our simulations, the models continue with the
processes of mixing, transfer of momentum and conversion of kinetic to
internal energy, however they seem to experience a kind of averaged
quasisteady evolution which can be still associated with the evolution
of a jet, i.e., a collimated flux of momentum. This jet is always
wider, slower and colder than the original one and is surrounded by a
broad shear layer. This Section is devoted to the examination of the
transition layers in distributions of gas density, jet mass fraction
and internal energy as well as shearing layers in velocity,
longitudinal momentum and Lorentz factor.

\begin{figure} 
\centerline{\psfig{file=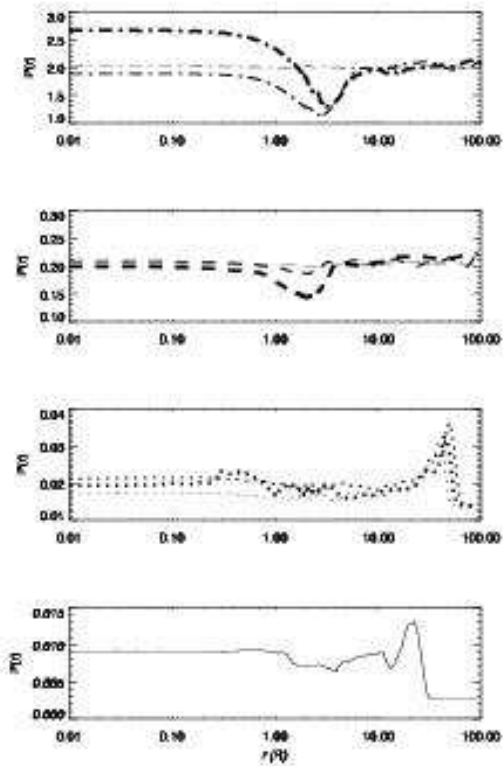,width=0.8\columnwidth}}
\caption{ Longitudinal averaged profiles of gas pressure for all
models.  Different types of lines are used for models with different
internal energies: Continuous line: model A; dotted line: model B;
dashed line: model C; dashed-dotted line: model D. Line thickness
increases with Lorentz factor (from 5, thinest line, to 20 thickest
one).  
}
\label{fig:press-profile}
\end{figure}

  Let us start by analyzing the overall structure of the pressure
field for the different models at the end of our
simulations. Figure~\ref{fig:press-profile} shows the transversal,
averaged profiles of pressure accross the computational grid. Several
comments are in order. In the case of models A and B, the shock formed
at the end of the saturation phase is seen propagating (at $r \approx
30 R_j$ in the case of model A, and at $r \approx 40-60 R_j$ in the
case of models B) pushed by the overpressure of the post shock state.
In the case of models C and D, the wave associated with the peak in the
pressure oscillation amplitude seem to have left the grid (remember
that in our jet models, hotter jets have also hotter ambient
media). The most remarkable feature in the pressure profile is the
depression centered at $r \approx 2 R_j$ in the case of models C10 and
C20 and at $r \approx 3 R_j$ in the case of models D10, D20. These
pressure minima coincides with the presence of vortices (clearly seen
in models C20 and D20 in the corresponding panels of
Fig.~\ref{fig:fbc10m} and Fig.~\ref{fig:fcd20m}). Also remarkable in
these plots is the almost total pressure equilibrium reached by models
C05 and D05 and the overpressure of the jet in model D20.

  As we noted in the previous Section the models evolve following four
schemes.  Jets belonging to the classes I and II disrupt leading to
the dispersion of tracer contours for more than five initial jet
radii. Model D20 of the II class is specific. It has not reached the
tracer contour dispersion equal to five jet radii, but it clearly
follows from Fig.~\ref{fig:tracerdisp} that this should happen around
$t=1100$. The models belonging to the classes III and IV do not
exhibit the dispersion of tracer contours for more than 5 jet radii
and look different at the end of simulations.

\begin{figure*}
\centerline{
\psfig{file=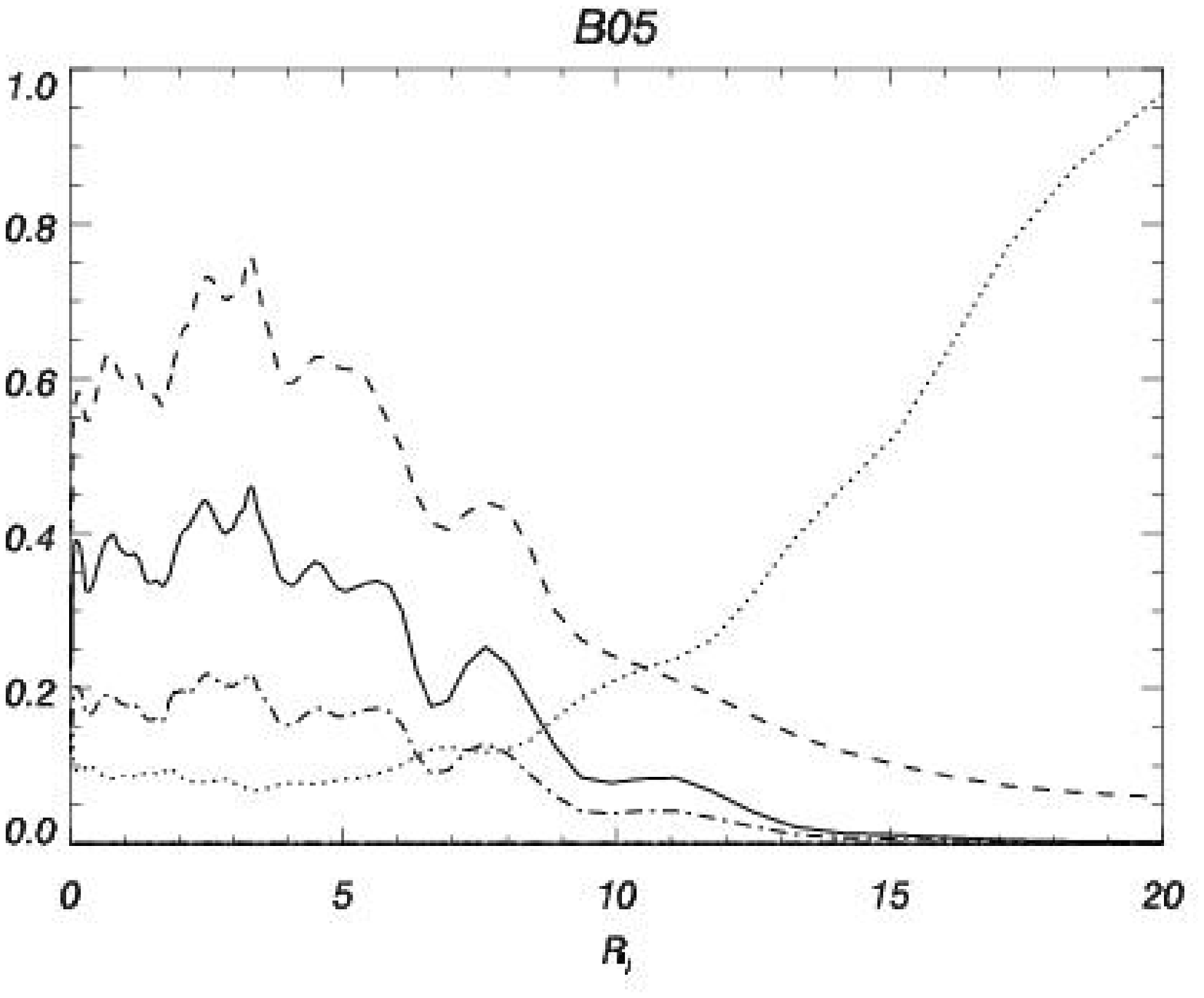,width=0.45\columnwidth} 
\psfig{file=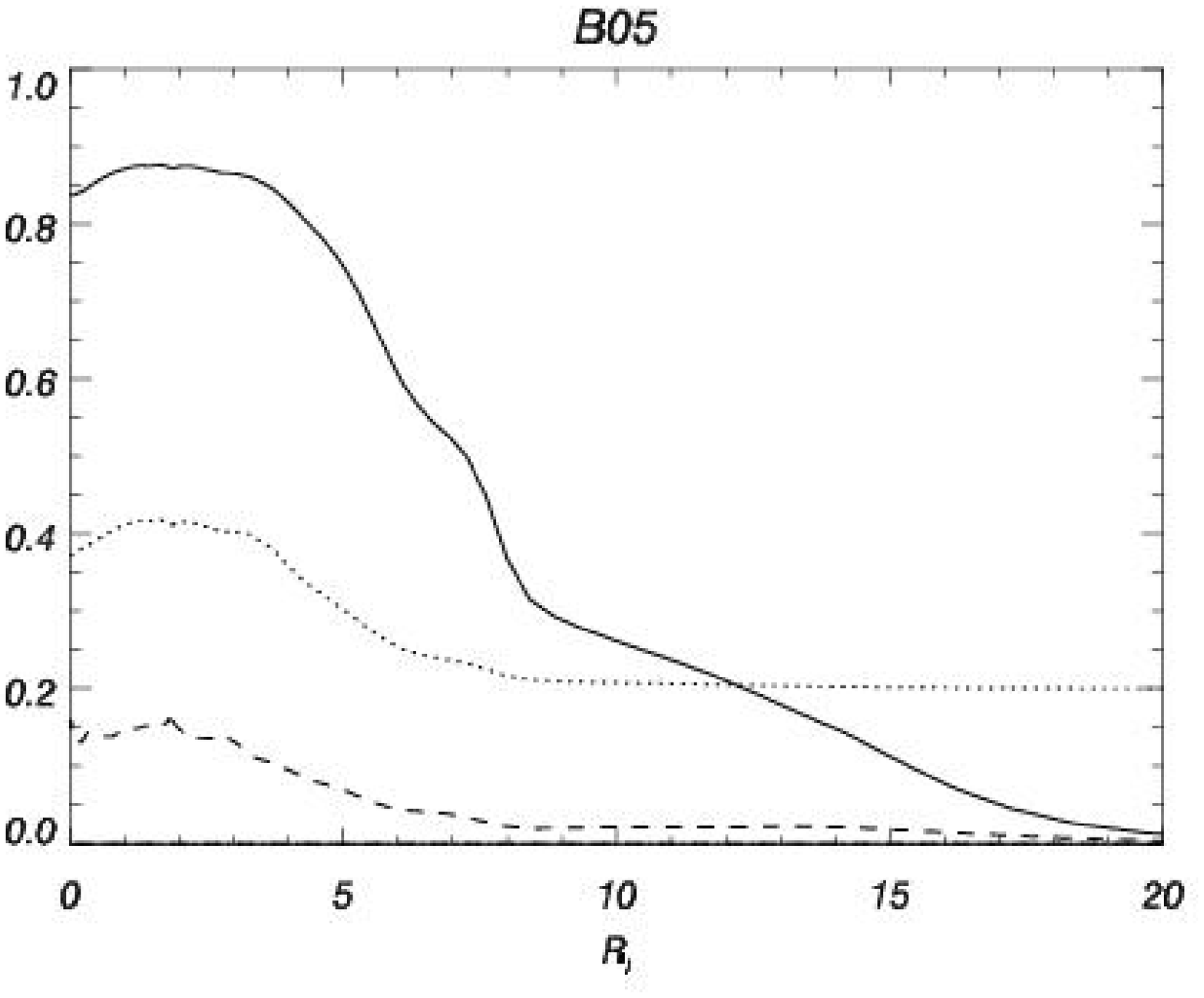,width=0.45\columnwidth} 
}
\centerline{
\psfig{file=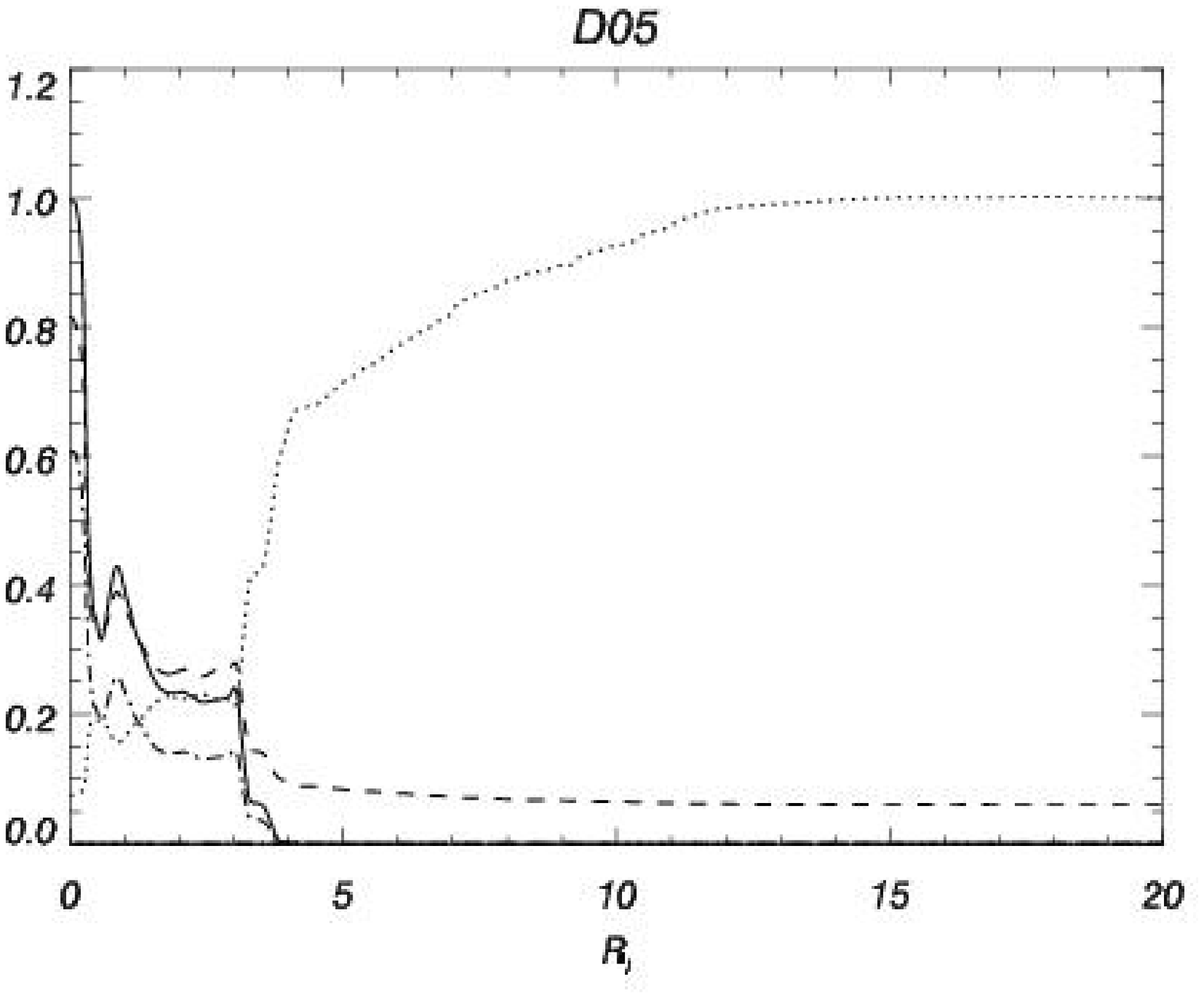,width=0.45\columnwidth} 
\psfig{file=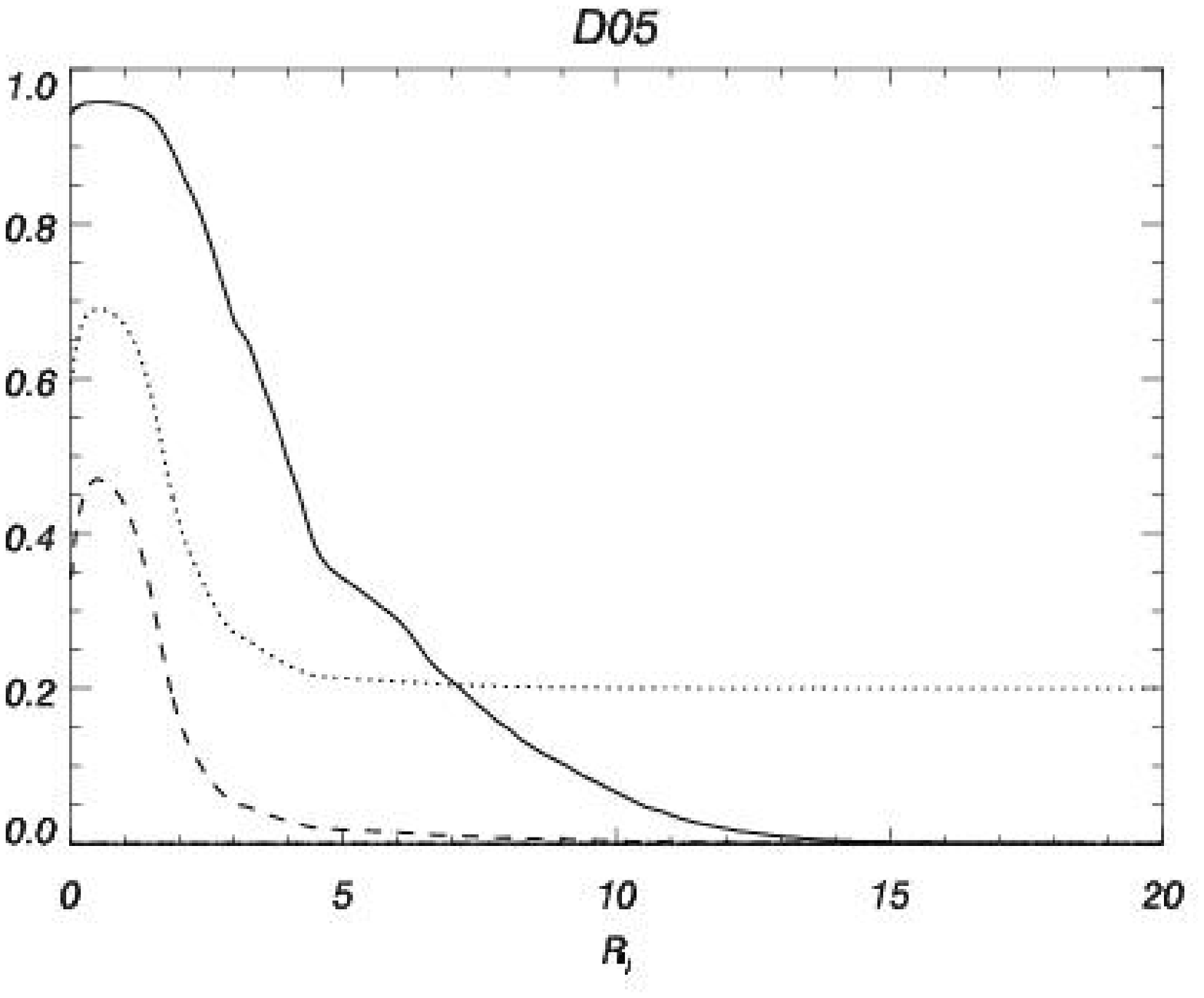,width=0.45\columnwidth} 
}

\caption{Transversal averaged profiles of relevant physical quantities
at the end of simulations B05 and D05. Left column: tracer, $f$ (full
line), rest mass density, $\rho_{0}$ (dotted line), specific internal
energy $\varepsilon$ (dashed line) and jet internal energy density , $e$
($=\rho_0\varepsilon f$; dash-dot line). Right column: longitudinal
velocity, $v_z$ (full line), Lorentz factor normalized to its initial
value, $\gamma/\gamma_0$ (doted line) and longitudinal momentum
normalized to its initial value, $S/S_0$. The upper plots represent
the model B05 and the lower plots represent D05. Note that the values
of $e$ are multiple by 10 for model B05 and divided by 10 for model
D05. The values of $\varepsilon$ for the model D05 are divided by 100.}

\label{fig:shear-profiles}
\end{figure*}

  Fig.~\ref{fig:shear-profiles} displays, for models B05 and D05
representing classes I/II and III/IV respectively, the profiles of
relevant physical quantities averaged along the jet at the end of the
simulations.  Let us note that different shear (in case of velocity
related quantities) or transition layers (in case of material
quantities) can be defined depending on the physical variable used.

  In case of model B05 all the material quantities (tracer, density
and internal energy) exhibit a wide broadening in the radial
direction. The distribution of tracer extends up to $r=15 R_j$, which
means that jet material has been spread up to this radius, with a
simultaneous entrainment of the ambient material into the jet
interior. The latter effect is indicated by the lowering of the
maximum tracer value from 1 down to 0.4. The fine structure of the
tracer distribution displays random variations, which apparently
correspond to the turbulent flow pattern well seen in
Fig.~\ref{fig:fcd05m} (upper set of panels). The curve of internal
energy is very similar to the one corresponding to tracer, however
variations are seen up to $r=20 R_j$. The profile of density is wider
than the profile of the tracer (the density is growing up to $r \sim 40
R_j$, which can be explained by the heating of external medium, in the
jet neighborhood, by shocks associated with outgoing large amplitude
sound waves and by transversal momentum transmitted to the ambient
medium via sound waves. Finally, the profile of the specific
internal energy is consistent with the density profile and the fact of
the jet being in almost pressure equilibrium.

  The dash-dot curve in Fig.~\ref{fig:shear-profiles} (top left panel)
represents the (internal) energy per unit volume held in jet matter.
Such a quantitiy, like the mean Lorentz factors in both inner jet and
shear layer are of special importance as they are directly related
to the emission properties of the model. Internal energy density in
jet particles is related to the fluid rest frame synchrotron
emissivity, whereas the fluid Lorentz factor governs the Doppler boosting
of the emitted radiation. As seen in the top right panel of
Fig.~\ref{fig:shear-profiles}, the final mean profile of velocity is
similar in shape to the profile of internal energy, despite the fact
that it is smoother. Similarly to internal energy, the longitudinal
velocity variations extend up to $r=20 R_j$. This can be understood in
terms of large amplitude, nonlinear sound waves, which contribute to
the transport of internal and kinetic energies in the direction
perpendicular to the jet axis. The profiles of Lorentz factor and
longitudinal momentum are significantly narrower.  Therefore in case
of models similar to B05 only the the most internal part, up to $r
\leq 8 R_j$ of the wide sheared jet, will be Doppler boosted, even
though the jet material quantities extend behind $r \simeq 15 R_j$.

  A similar discussion can be performed for the model D05 representing
the other group of jets, which form a shear layer without experiencing
the phase of rapid disruption. As seen in the bottom panels of
Fig.~\ref{fig:shear-profiles}, the jet of model D05 preserves sharp
boundaries between their interior and the ambient medium, although
both media are modified by the dynamical evolution at earlier
phases. The sharp boundary (transition layer) at $r\simeq 3.2 R_j$ is
apparent in the plots of all material quantities, i.e. tracer,
density, internal energy. The thickness of the transition layer for
all the quantities is comparable to one initial jet radius, 10 times
narrower than in case of B05.  We note, however a smooth change of
ambient gas density in the range of $r\sim 3.2 R_j \div 12 R_j$. 

  It is apparent also that a narrow core of almost unmixed ($f=1$) jet
material remains at the center in the currently discussed case. The
radius of the core is about one half of the original jet radius. The
core sticks out from a partially mixed, relatively uniform sheath and
is well seen in the plots of tracer and internal energy for that
model, however it disappears when increasing the resolution in the
longitudinal direction, as seen in the Appendix.

  Concerning the dynamical quantities, we note that there is no sharp
jump in the profiles of longitudinal velocity, Lorentz factor and
longitudinal momentum and the central core does not appear in profiles
of these quantities. Significant longitudinal velocities extend up to
$r \simeq 12 R_j$ as in case of density, in contrast to tracer and
internal energy. As noted previously, the averaged pressure
distribution for model D05 is practically uniform in the whole
presented range of the transversal coordinate. Therefore as in case
B05 we can conclude that the variations of density in the ambient
medium are due to the heat deposited by nonlinear sound waves. On the
other hand the widths of the profiles of the Lorentz factor and
longitudinal momentum are comparable to those of jet mass fraction and
specific internal energy. Then the emission of the whole jet volume
will be Doppler boosted.

  Models B05 and D05 were considered as representative cases of models
developing shear layers wider (group 1; classes I and II) and narrower
(group 2; classes III and IV), respectively, than $5 R_j$. Now the
question is up to which extent the characteristics of the shearing
flow of these two models are common to the models in the corresponding
groups. We note that given the large differences between the
initial parameters of models in classes I and II, on one hand, and III
and IV, on the other, we do not expect a perfect match among the
properties of the transversal structure in models within the same
group. For example, whereas models in class I develop wide shear
layers due to the action of a strong shock formed at the end of the
saturation phase, models in class II develop shear layers through a
continuous injection of transversal momentum and the generation of
large vortices at the jet ambient interface.

  We now investigate relations among the following averaged quantities
in the whole set of models at the final state: the dispersion of
tracer contours, the typical widths of profiles of density, internal
energy density in jet matter ($r_e$), velocity, Lorentz factor
($r_\gamma$) and longitudinal momentum ($r_S$) and the peak values of
Lorentz factor and the longitudinal momentum ($S_{\rm max}$). In all
cases the peak values were measured directly, whereas the typical
width of the profiles were taken as their width at the mean value
between the maximum and minimum ones. We find that for all models of
group 1, $r_{\gamma}>4 R_j$, $r_S > 3.5 R_j$, $r_e > 7R_j$ and $S/S_0
< 0.2$. In case of all models of group 2, $r_{\gamma} <4 R_j$, $r_S <
3.5 R_j$, $r_e < 7 R_j$ and $S_{\rm max}/S_0 > 0.2$. 

\section{Discusion and conclusions \label{sect:concl}}

  We have studied the non-linear evolution of the relativistic planar
jet models considered in Paper I. The initial conditions considered
cover three different values of the jet Lorentz factor (5, 10, 20) and
a few different values of the jet specific internal energy (from
0.08$c^2$ to 60.0$c^2$). The models have been classified into four
classes (I to IV) with regard to their evolution in the nonlinear
phase, characterized by the process of mixing and momentum
transfer. Cold, slow jets (Class I) develop a strong shock in the
jet/ambient interface at the end of the saturation phase leading to
the development of wide, mixed, shear layers. Hot fast models (Class
II) develop wide shear layers formed by distinct vortices and transfer
more than 50\% of the longitudinal momentum to the ambient medium. In models
within this class, the high Lorentz factor in the original jet and its
high internal energy act as a source of transversal momentum that
drives the process of mixing and momentum transfer. Between these
classes we find hot, slow models (Class III) that have intermediate
properties. Finally we have found warm and fast models (Class IV) as
the most stable. Whether a jet is going to develop a strong shock and
be suddenly disrupted seems to be encoded in the peak of the pressure
oscillation amplitude at the end of the saturation phase and the
related transversal Mach number.

  The above picture is clarifying but is subject to the limitations of
our choice of initial parameters that was restricted to values with
$\rho_{\rm 0j} = 0.1$ (see Paper I). This restriction together with
the initial pressure equilibrium lead to a constant jet-to-ambient
ratio of specific internal energies for all the models, i.e., hotter
jets are surrounded by hotter ambient media. In order to extend our
conclusions to a wider region in the initial parameter space, we have
performed a supplementary set of simulations (F-L) with the aim of
disentangling the effect of the ambient medium in the development of
the disruptive shock appearing after saturation. Thus, 
hot, tenuous, slow/moderately fast
jets (F, G, H, I, L) behave like cold, dense ones in a cold
environment (A05, B05, B10). However, if these hot, tenuous jets are
faster (J, K), they behave as warm, fast models (e.g., C10, C20,
B20). The fact that the initial Lorentz factor is high seems to prevent
the transversal velocity from growing enough to generate the strong shock
which breaks the slower jets.

\begin{figure} 
\centerline{
\psfig{file=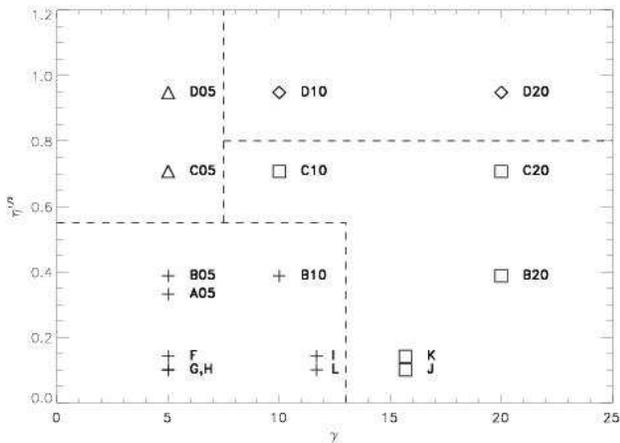,width=\columnwidth}}
\caption{Square root of the jet-to-ambient enthalpy ratio (see Paper I
  for definitions) versus jet Lorentz factor. Symbols represent
  different non-linear behaviors: crosses stand for shock disrupted
  jets (cold, slow jets, along with tenuous, hot, moderately fast or
  slow ones); diamonds for unstable, hot, fast jets; triangles for
  relatively stable hot, slow, and squares for stable, warm, fast,
  along with hot, tenuous, faster jets.}
\label{fig:stabplane}
\end{figure}

  Models undergoing qualitatively
different non-linear evolution are clearly grouped in well-separated
regions in a jet Lorentz factor/jet-to-ambient enthalpy diagram (see
Fig.~\ref{fig:stabplane}). Models in the lower, left corner (low
Lorentz factor and small enthalpy ratio) are those disrupted by a
strong shock after saturation. Those models in the upper, left corner
(small Lorentz factor and hot) represent a relatively stable region.
Those in the upper right corner (large Lorentz factor and enthalpy ratio)
are unstable although the process of mixing and momentum exchange
proceeds on a longer time scale due to a steady conversion
of kinetic to internal energy in the jet. Finally, those in the lower,
right region (cold/warm, tenuous, fast) are stable in the nonlinear
regime.

  Our results differ from those of Mart\'{\i} et al. (1997), Hardee et
al. (1998) and Rosen et al. (1999) who found fast, hot jets as the
more stable. The explanation given by Hardee et al. (1998) invoking
the lack of appropriate perturbations to couple to the unstable modes
could be partially true as fast, hot jets do not generate
overpressured cocoons that let the jet run directly into the
nonlinear regime.  However, as pointed out in Paper I, the high
stability of hot jets may have been caused by the lack of radial
resolution, that leads to a damping in the perturbation growth
rates. Finally, the simulations performed
in the aforementioned papers only covered about one hundred time
units, well inside the linear regime of the corresponding models for
small perturbations. In this paper, the problem of the stability of
relativistic jets is analyzed on the basis of long-term simulations
that extend over the fully nonlinear evolution of KH instabilities.

  At the end of our simulations, the models continue with the
processes of mixing, transfer of momentum and conversion of kinetic to
internal energy, however they seem to experience a kind of averaged
quasisteady evolution which can be still associated with the evolution
of a jet, i.e., a collimated flux of momentum. This jet is always
wider, slower and colder than the original one and is surrounded by a
distinct shear layer. Hence transversal jet structure naturally
appears as a consequence of KH perturbation growth. The widths of
these shear (in case of velocity related quantities) or transition
layers (in case of material quantities) depend on the specific
parameters of the original jet model as well as the physical variable
considered. However, models in classes III and
IV develop thin shear layers, whereas the shear layers of models in
classes I and II are wider. The possible connection of these results
to the origin of the FRI/FRII morphological dichotomy of jets in
extended radiosources will be the subject of further
research. Extensions of the present study to models with a
superposition of perturbations, cylindrical symmetry and three
dimensions are currently underway.

\begin{acknowledgements}
  The authors want to thank H. Sol for clarifying discussions during
earlier phases of development of this work.  This work was supported
in part by the Spanish Direcci\'on General de Ense\~nanza Superior
under grant AYA-2001-3490-C02 and by the Polish Committee for
Scientific Research (KBN) under grant PB
404/P03/2001/20. M.H. acknowledges financial support from the visitor
program of the Universidad de Valencia. M.P. has benefited from a
predoctoral fellowship of the Universidad de Valencia ({\it V Segles}
program).
\end{acknowledgements}
\begin{appendix}
\section{Influence of numerical resolution in the nonlinear
          evolution}

  In order to be aware of the limitations of the resolution used in
the results for the non-linear regime, we repeated model C05 (C16
here) with the same transversal resolution (400 cells/$R_{\rm j}$) and
changing the longitudinal resolution. First we double it (32
cells$/R_{\rm j}$, C32) and then multiply it by four (64 cells$/R_{\rm
j}$,C64), and finally, in order to have similar resolutions in both 
transversal and longitudinal directions, we performed this simulation
using 256$\times$128 cells$/R_{\rm j}$ (C128). Models C16, C32 and C64
were evolved up to a time larger than 600 $R_{\rm j}/c$, whereas model
C128 was stopped at $t=375 R_{\rm j}/c$.

  Table~\ref{tab:res1} displays the data corresponding to
Table~\ref{tab:phases} for models C16, C32, C64, C128. Differences in
the duration of the phases are apparent but not significant. Regarding
the linear regime, perturbations in models C32 and C64 grow closer to
linear predictions (growth rate 0.093 $c/R_{\rm j}$) in both cases; to
be compared with the analytical value 0.114 $c/R_{\rm j}$) rather 
than C16 (0.085 $c/R_{\rm j}$). C128 has a slower growth rate (0.073
$c/R_{\rm j}$) due to its smaller transversal resolution (see Appendix
of Paper I). Values of pressure perturbation at the peak range from 3
to 8, increasing generally from smaller to larger longitudinal resolution.
Moreover, transversal relativistic Mach numbers are also increasing
with resolution (from a value of about unity to two), so that we
observe a stronger, although still weak, shock in C64 and C128 than in
C16 or C32. This is maybe due either to the instability giving
rise to the shock (see Section \ref{ss:mixdis}) being better captured
with increasing resolution or as the result of a smaller numerical
viscosity.

\begin{figure*}
\centerline{ \psfig{file=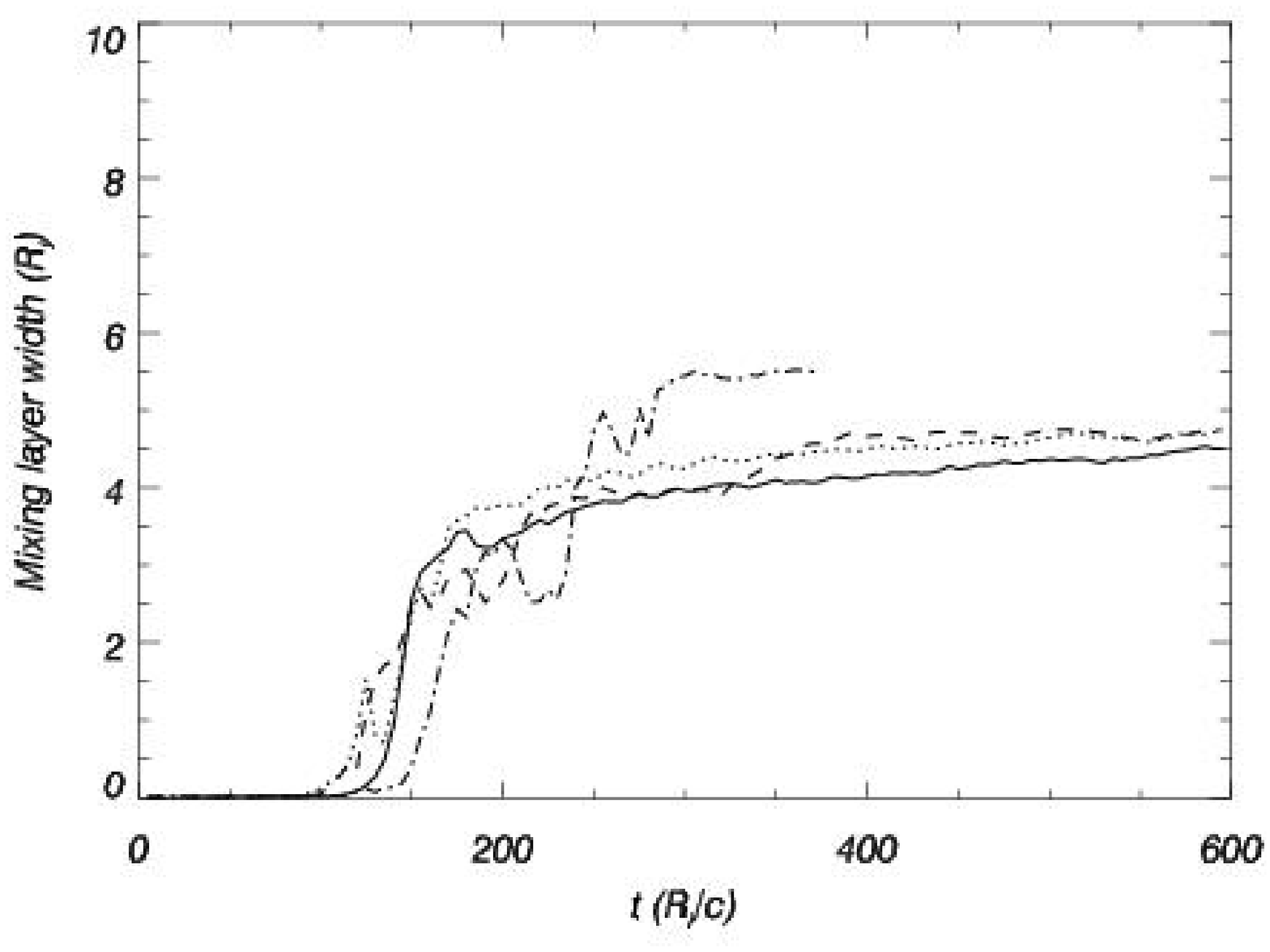,width=0.45\columnwidth}
\psfig{file=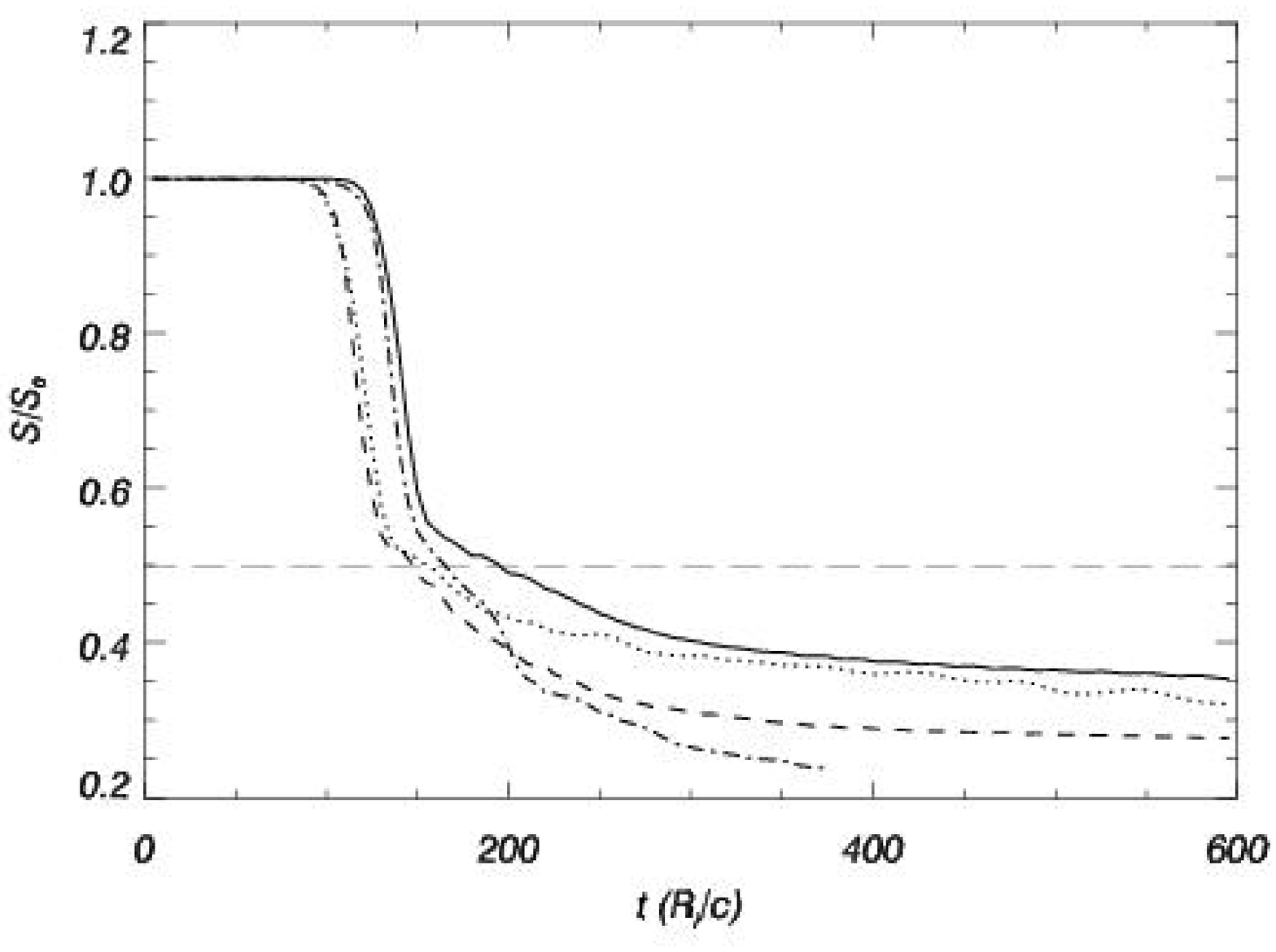,width=0.45\columnwidth} } 
\caption{Same plots
as \ref{fig:tracerdisp} (left) and \ref{fig:axialm} (right) for models C16
(solid line), C32 (dotted line), C64 (dashed line) and C128
(dash-dotted line).
} 
\label{fig:Smix}
\end{figure*}

  Figure~\ref{fig:Smix} shows the time evolution of the mean width of
the jet/ambient mixing layer and the total longitudinal momentum in
the jet for model C05 as a function of resolution. No noticeable
differences are found in the evolution of the different numerical
simulations within the linear phase (up to $t\approx 100 R_{\rm
j}/c$). However, there is a clear tendency to develop wider mixing
layers and enhance momentum transfer in those simulations with higher
numerical resolutions as a result of the reduction of numerical
viscosity. In the case of model C128 the processes of jet/ambient
mixing and momentum exchange are further enhanced by the ratio of
longitudinal to transversal resolution close to unity which favors the
generation of vortices in the jet/ambient boundary. The enhancement of
mixing with numerical resolution and the generation of vortices in
model C128 is seen in the sequence of
Panels for different simulations (see published paper) 
show that models C16 and C32 are very similar, whereas models C64 and C128
are totally mixed (see the maxima of the tracer values in the scales)
and colder (as a result of the enhanced mixing with the cold ambient
medium).

  Finally, mean transversal profiles (appearing in the published paper) of
relevant physical quantities show that the thin, hot core in model 
C16 disappears in model C128
due to the enhanced mixing. In model C32 mixing down to the axis
occurs after a long process. Models C64 and C128 develop weak shocks
after saturation and suffer sudden mixing, which goes on in time,
cooling down the remaining relativistic flow. Transition layers in
rest mass density, jet mass fraction and specific internal energy are
wider in model C128 due to the enhanced mixing. Longitudinal velocity
and momentum and Lorentz factor profiles are more similar in all the
cases although the relativistic core is thinner in model C128.

\begin{table*}
$
\begin{array}{cccccccc}
\hline
{\rm Model}   & t_{\rm lin} & t_{\rm mix} & t_{\rm sat} & t_{\rm peak}
& \Delta_{\rm peak} & t_{\rm fmix}  & t_{\rm meq} \\
\hline
{\rm C16}  & 100 & 120   &  125  & 130 & 5.0  & >595   & 195 \\
{\rm C32}  & 85  & 100   &  105  & 105 & 3.5  &  500   & 150 \\
{\rm C64}  & 80  & 100   &  110  & 115 & 6.5  &  125   & 150 \\
{\rm C128} & 100 & 125   &  110  & 125 & 8.0  &  135   & 160 \\
\hline
\end{array}
$ 
\caption{Times for different phases in the evolution of simulations
C16, C32, C64 and C128. See Table \ref{tab:phases} for the meaning of
the times at the table entries. }
\label{tab:res1}
\end{table*}
\end{appendix}


\end{document}